\documentclass[useAMS,usenatbib,usegraphicx]{mn2e}

\usepackage{url}
\usepackage[varg]{txfonts}
\usepackage{subfigure}
\usepackage{times}
\usepackage{amssymb}
\usepackage{xspace}
\def\uvic{Dept. of Physics and Astronomy, 
  University of Victoria, Victoria, BC, V8P 5C2, Canada}
\def\lick{UCO/Lick Observatory, Department of Astronomy and Astrophysics, 
  University of California, Santa Cruz, CA 95064, USA}
\def\ucsb{Dept. of Physics, University of California, 
  Santa Barbara, CA 93106, USA}
\def\kipac{Kavli Institute for Particle Astrophysics and Cosmology, 
 Stanford University, 452 Lomita Mall, Stanford, CA 94035, USA}
\def\utah{Department of Physics and Astronomy, University of Utah, 
  Salt Lake City, UT 84112, USA}
\def\kapteyn{Kapteyn Astronomical Institute, University of Groningen, 
  P.O.Box 800, 9700 AV Groningen, The Netherlands}
\def\oxford{Department of Physics, University of Oxford, 
  Keble Road, Oxford, OX1 3RH, UK}
\def\cambridge{Institute of Astronomy, University of Cambridge,
  Madingley Rd, Cambridge, CB3 0HA, UK}

\def\barnabeemail{\tt mbarnabe@stanford.edu}

\def\packard{Packard Research Fellow}
\def\cita{CITA National Fellow}

\newcommand{\aj}{AJ}       % Astronomical Journal
\newcommand{\apj}{ApJ}     % Astrophysical Journal
\newcommand{\apjl}{ApJ}    % Astrophysical Journal Letters
\newcommand{\mnras}{MNRAS} % Monthly Notices of the Royal Astronomical Society
\newcommand{\aap}{A\&A}    % Astronomy and Astrophysics
\newcommand{\araa}{ARAA}   % Annual Reviews in Astronomy and Astrophysics
\newcommand{\pasp}{PASP}   % Publications of the Astr. Society of the Pacific
\newcommand{\pasj}{PASJ}   % Publications of the Astr. Society of Japan

\newcommand{\nat} {Nature}

\def\Sref#1{Section~\ref{#1}\xspace}
\def\Fref#1{Figure~\ref{#1}\xspace}
\def\Tref#1{Table~\ref{#1}\xspace}

\def\Eqref#1{Eq.~(\ref{#1})\xspace}

\renewcommand\vec[1]{\bmath{#1}}
\newcommand\mat[1]{\mathbf{#1}}

\newcommand {\kms} {\ifmmode  \,\rm km\,s^{-1} \else $\,\rm km\,s^{-1}  $ \fi }
\newcommand {\kpc} {\ifmmode  {\rm kpc}  \else ${\rm  kpc}$ \fi  }  
\newcommand {\pc} {\ifmmode  {\rm pc}  \else ${\rm pc}$ \fi  }  
\newcommand {\Msun} {\ifmmode {\rm M_{\odot}} \else ${\rm M_{\odot}}$ \fi} 
\newcommand {\Zsun} {\ifmmode {\rm Z_{\odot}} \else ${\rm Z_{\odot}}$ \fi} 
\newcommand {\yr} {\ifmmode yr^{-1} \else $yr^{-1}$ \fi} 
\newcommand {\hMsun} {\ifmmode h^{-1}\,\rm M_{\odot} \else $h^{-1}\,\rm M_{\odot}$ \fi}

\usepackage[usenames]{color}

\newcommand{\vc}{v_{\rm c}}

\newcommand{\veceta}{\vec{\eta}}

\newcommand{\de}{\mathrm{d}}
\newcommand{\zL}{z_{\mathrm{lens}}}

\newcommand{\cauldron}{\textsc{cauldron}}

\newcommand{\jdisk}{SDSS\,J2141\xspace}

\newcommand{\slope}{\gamma}

\newcommand{\REin}{R_{\mathrm{E}}}

\newcommand{\Jz}{J_{z}}

\newcommand{\fDM}{f_{\mathrm{DM}}}
\newcommand{\MDM}{M_{\mathrm{DM}}}
\newcommand{\Mtot}{M_{\mathrm{tot}}}

\newcommand{\pr}{\mathrm{Pr}\,}
\newcommand{\Phit}{\Phi_{\mathrm{tot}}}

\newcommand{\Phistar}{\Phi_{\star}}
\newcommand{\PhiDM}{\Phi_{\mathrm{DM}}}
\newcommand{\bvRq}{\overline{v_{R}^{2}}}
\newcommand{\bvfq}{\overline{v_{\varphi}^{2}}}

\newcommand{\bvf}{\overline{v_{\varphi}}}
\newcommand{\bvzq}{\overline{v_{z}^{2}}}
\newcommand{\bvRvz}{\overline{v_{R} v_{z}}}
\newcommand{\sigfq}{\sigma^{2}_{\varphi}}

\newcommand{\rhoDM}{\rho_{\mathrm{DM}}}
\newcommand{\rhostar}{\rho_{\star}}
\newcommand{\MoL}{\Upsilon}

\newcommand{\MoLs}{\Upsilon_{\star}}
\newcommand{\Ms}{M_{\star}}
\newcommand{\rs}{r_{\rm s}} %% scale radius
\newcommand{\rsg}{r_{-2}} 
\newcommand{\rvir}{r_{\mathrm{vir}}} %% virial radius
\newcommand{\vvir}{v_{\mathrm{vir}}} %% virial velocity
\newcommand{\Mvir}{M_{\mathrm{vir}}} %% virial mass
\newcommand{\vbar}{v_{\mathrm{bar}}}   %% v baryonic
\newcommand{\vdisk}{v_{\mathrm{disk}}} %% v disk
\newcommand{\vtot}{v_{\mathrm{tot}}} %% v total
\newcommand{\cg}{c_{-2}} 
\newcommand{\qh}{q_{\rm h}}
\newcommand{\Dd}{D_{\mathrm{d}}}
\newcommand{\Ds}{D_{\mathrm{s}}}
\newcommand{\Dds}{D_{\mathrm{ds}}}
\newcommand{\xt}{\tilde{x}}
\newcommand{\yt}{\tilde{y}}
\newcommand{\Scr}{\Sigma_{\mathrm{crit}}}
\newcommand{\Rd}{R_{\mathrm{d}}} %% disk scale radius
\newcommand{\orb}{\sigma_{R}^{2}/\sigma_{z}^{2}}

\title[The SWELLS survey. IV]{The SWELLS survey. IV.  Precision
  measurements of the stellar and dark matter distributions in a
  spiral lens galaxy}

\author[Barnab\`e et al.]{% 
  Matteo~Barnab\`e,$^{1}$\thanks{\barnabeemail} 
  Aaron~A.~Dutton,$^{2,3,4}$\thanks{\cita}
  Philip~J.~Marshall,$^{5}$
  Matthew~W.~Auger,$^{2,6}$
\newauthor{%
  Brendon~J.~Brewer,$^{2}$
  Tommaso~Treu,$^{2}$\thanks{\packard}
  Adam~S.~Bolton,$^{7}$
  David~C.~Koo$^{3}$}
\newauthor{%
  and L\'eon~V.~E.~Koopmans$^8$}\\
  $^1$\kipac\\
  $^2$\ucsb\\
  $^3$\lick\\
  $^4$\uvic\\
  $^5$\oxford\\ 
  $^6$\cambridge\\
  $^7$\utah\\
  $^8$\kapteyn\\
}

\begin{document}

\date{}

\maketitle

\label{firstpage}

\begin{abstract} 
We construct a fully self-consistent mass model for the lens galaxy
{\jdisk} at redshift $0.14$, and use it to improve on previous studies
by modelling its gravitational lensing effect, gas rotation curve and
stellar kinematics simultaneously.  We adopt a very flexible
axisymmetric mass model constituted by a generalized NFW dark matter
halo and a stellar mass distribution obtained by deprojecting the
multi-Gaussian expansion (MGE) fit to the high-resolution K'-band
Laser Guide Star Adaptive Optics (LGSAO) imaging data of the galaxy,
with the (spatially constant) mass-to-light ratio as a free
parameter. We model the stellar kinematics by solving the anisotropic
Jeans equations.  We find that the inner logarithmic slope of the dark
halo is weakly constrained, i.e. $\slope = 0.82^{+0.65}_{-0.54}$, and
consistent with an unmodified NFW profile; we can conclude, however,
that steep profiles ($\slope \ge 1.5$) are disfavoured ($<14$\%
posterior probability). We marginalize over this parameter to infer
the galaxy to have (i) a dark matter fraction within 2.2 disk radii of
$0.28^{+0.15}_{-0.10}$, independent of the galaxy stellar population,
implying a maximal disk for {\jdisk}; (ii) an apparently uncontracted
dark matter halo, with concentration $\cg = 7.7_{-2.5}^{+4.2}$ and
virial velocity $\vvir = 242_{-39}^{+44}$ $\kms$, consistent with
$\Lambda$CDM predictions; (iii) a slightly oblate halo ($q_h =
0.75^{+0.27}_{-0.16}$), consistent with predictions from
baryon-affected models.  Comparing the tightly constrained
gravitational stellar mass inferred from the combined analysis
($\log_{10} \Ms/\Msun = 11.12_{-0.09}^{+0.05}$) with that inferred
from stellar populations modelling of the galaxies colours, and
accounting for an expected cold gas fraction of $20 \pm 10$ per cent,
we determine a preference for a Chabrier IMF over Salpeter IMF by a
Bayes factor of 5.7 (corresponding to substantial evidence).  We infer
a value $\beta_{z} \equiv 1 - \sigma^{2}_{z}/\sigma^{2}_{R} =
0.43_{-0.11}^{+0.08}$ for the orbital anisotropy parameter in the
meridional plane, in agreement with most studies of local disk
galaxies, and ruling out at 99~per cent confidence level that the
dynamics of this system can be described by a two-integral
distribution function.
\end{abstract}

\begin{keywords}
  gravitational lensing: strong     ---
  galaxies: fundamental parameters  ---
  galaxies: haloes                  ---
  galaxies: kinematics and dynamics ---
  galaxies: spiral                  ---
  galaxies: structure               
\end{keywords}

%%%%%%%%%%%%%%%%%%%%%%%%%% INTRODUCTION %%%%%%%%%%%%%%%%%%%%%%%%%%%%%% 

\clearpage
\newpage

\section{Introduction}
\label{sec:introduction}

Measuring the relative contribution of luminous and dark matter in
spiral galaxies is essential to understand their internal structure
and therefore constrain the physical processes that drive their
formation and evolution \citep[e.g.][]{Dutton2011-DH}. Traditionally
this is done by means of detailed stellar and gas kinematics and
stellar population diagnostics \citep[e.g.][]{Bershady2011}. However,
often one needs additional assumptions about the relative contribution
of the stars and dark matter \citep[e.g.][]{vanAlbada-Sancisi1986}, or
about the stellar initial mass function
\citep[e.g.][]{Bell-deJong2001}.

The combination of strong gravitational lensing and galaxy kinematics
is a powerful tool for constraining the mass distribution and the
dynamical structure of galaxies beyond the local Universe, since this
approach makes it possible to overcome many of the difficulties
associated with the traditional techniques, which are severely limited
when applied to distant objects \citep[see e.g.][]{Treu-Koopmans2002a,
  Treu-Koopmans2004, Barnabe-Koopmans2007, Jiang-Kochanek2007,
  Grillo2008, Auger2010, Koopmans2009, Barnabe2010, Barnabe2011}. In
particular, gravitational lenses in which the deflector contains a
high inclination disk provide extra (geometrical) information to help
disentangle the distribution of baryons and dark matter, and to
measure the 3D shape of the dark matter halo (e.g.,
\citealt{Keeton-Kochanek1998}, \citealt{Koopmans1998},
\citealt{Maller2000}, \citealt{Trott2010}, \citealt{Suyu2011} and
\citealt{Dutton2011-SW}, Paper~II of this series) Because these
measured masses are gravitational, they can be compared with the
stellar mass from stellar population synthesis (SPS) models and so
used to constrain the form of the stellar initial mass function (IMF),
and the response of dark matter haloes to galaxy formation.

Until recently only a small number of gravitational lenses with
high-inclination disks were known.  The SLACS \citep{Bolton2006,
  Bolton2008a} and SWELLS \citep[][Paper~I]{Treu2011} surveys have
significantly increased the number of known gravitational lenses in
which the deflector contains a high-inclination disk, including
several disk-dominated systems.

One of the most promising spiral lens systems for a joint lensing and
dynamics analysis is SDSS\,J2141$-$0001 (hereafter simply referred to
as {\jdisk} for brevity), at redshift $\zL = 0.1380$, which is a
disk-dominated galaxy (it has a disk K'-band light fraction of $\simeq
80\%$) at high inclination ($i\simeq78^{\circ}$).  In addition to the
discovery data from the SLACS survey, a wealth of imaging and
kinematic data are available from the SWELLS project (Paper~I
and~II). A joint strong lensing and gas kinematics (rotation curve)
analysis of {\jdisk} conducted in Paper~II yielded a gravitational
stellar mass of $\log_{10} (\Ms/\Msun) = 10.99^{+0.11}_{-0.25}$
(consistent with that from a stellar population analysis assuming a
Chabrier IMF), a dark matter fraction at 2.2 disk scale lengths of
$\fDM = 0.55^{+0.20}_{-0.15}$, and a dark matter halo flattening of
$\qh = 0.91^{+0.15}_{-0.13}$.  However, in that work, simple
phenomenological (``Chameleon'') models for all three mass components,
i.e. the dark matter halo and the stellar bulge and disk, were
assumed.  Moreover, only a fraction of the available kinematic data
for the lens galaxy was used: the stellar velocity dispersion and
rotation curve were not considered. Indeed, the velocity dispersion
could not be predicted self-consistently within the assumed model.

In this paper we improve on the Paper~II analysis of {\jdisk} in
several important ways. The main improvement is the inclusion of
stellar kinematics data, which provides a mass constraint at smaller
radii than obtained from lensing or gas kinematics. It is well known
that disk galaxies are usually characterized by a velocity dispersion
ellipsoid flattened along the vertical direction: therefore, in order
to provide an accurate description of the data set, we model the
stellar kinematics by means of anisotropic Jeans equations, which
allow us to properly take into account (and recover) the anisotropy
ratio parameter $\beta_{z}$. In addition to this, we use more flexible
and general models for both the stellar and the dark matter density
profiles.  Specifically, we obtain the stellar mass density profile
from the deprojection of the observed luminous distribution (fitted
with the state-of-the-art method of multi-Gaussian expansion, MGE, see
\citealt{Cappellari2002}), rather than the sum of two ``Chameleon''
profiles, which were used to approximate a S\'ersic profile bulge
\citep{Sersic1968} and an exponential disk. Finally, here we model the
dark matter halo with an ellipsoidal generalized NFW profile (inner
logarithmic slope $-\gamma$, outer slope $-3$) rather than the
non-singular isothermal ellipsoid (inner slope $0$, outer slope $-2$)
with a fixed density profile in the inner regions used in the previous
analysis.

The resulting model is both self-consistent and, in the case of the
dark matter halo, physically-motivated, and allows us to attempt to
fit all the data we have for {\jdisk} simultaneously. We use it to
answer the following questions about {\jdisk}: How much does dark matter
contribute to the total mass of this disk galaxy, in particular in the
inner regions? What is the concentration and inner profile slope of
its dark matter halo?  What is its halo's shape? When calibrated via
its stellar mass distribution's gravitational effects, what
galaxy-averaged IMF do we infer from a stellar population synthesis
analysis of its optical and near infra-red colours? What is the
vertical-to-radial anisotropy of its velocity dispersion ellipsoid?

This paper is structured as follows. We first describe our
observational data (imaging for the lens modelling, spectroscopy for
the stellar and gas kinematics) in \Sref{sec:observations}. We then
outline our mass model for {\jdisk} in \Sref{sec:mass-model}, giving
the functional forms we use to describe its stellar and dark matter
distributions. Then, in \Sref{sec:lensing} and \Sref{sec:kinematics}
we show how our model predicts both the lensing and kinematic data in
a self-consistent way, and in \Sref{sec:bayes} we review the
probability theory behind the actual inference procedure we follow.
In \Sref{sec:results} we present and discuss the results of our
analysis, and in \Sref{sec:conclude} we draw conclusions, providing an
answer to the questions posed above.

Throughout this work, we assume a flat $\Lambda$CDM cosmology with
present day matter density, $\Omega_{\rm m}=0.3$, and Hubble
parameter, $H_0=70 \rm\,km\,s^{-1}\,Mpc$.

%%%%%%%%%%%%%%%%%%%%%%%%%%% OBSERVATIONS %%%%%%%%%%%%%%%%%%%%%%%%%%%%% 

% --------------- BEST RECONSTRUCTION: LEN --------------------------
\begin{figure}
  \centering
  \resizebox{1.00\hsize}{!}{\includegraphics[angle=-90]
            {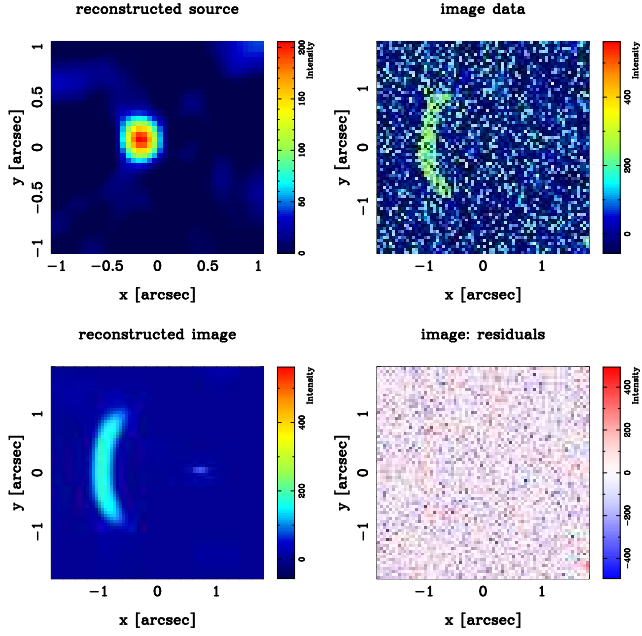}}
  \caption{Lensed image reconstruction obtained from the \emph{maximum
      a posteriori} model. From the top left-hand to bottom right-hand
    panel: reconstructed source; \textit{HST}/ACS data showing the
    observed lensed image after subtraction of the lens galaxy; lensed
    image reconstruction corresponding to the source in the first
    panel; residuals.}
  \label{fig:LENcomp}
\end{figure}
% -------------------------------------------------------------------

% --------------- BEST RECONSTRUCTION: JAM --------------------------
\begin{figure}
  \centering
  \resizebox{1.00\hsize}{!}{\includegraphics[angle=-90]
            {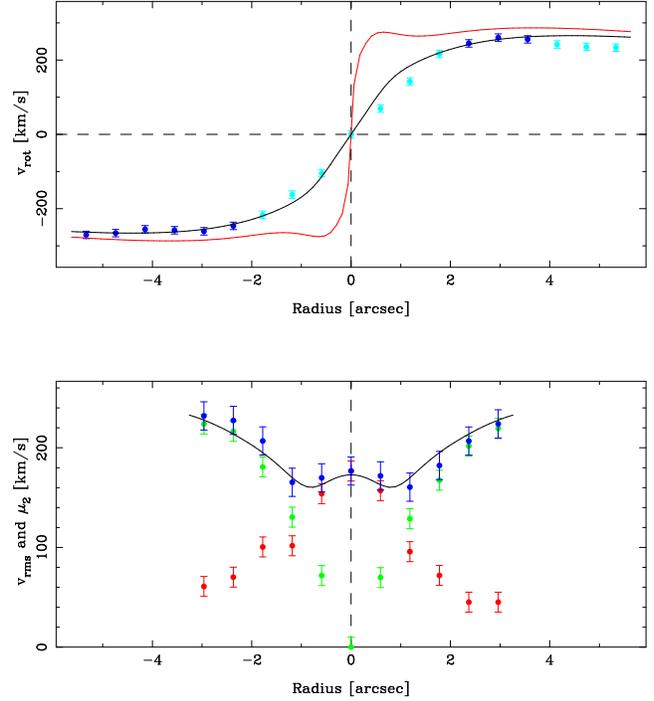}}
  \caption{Observed gas and stellar kinematics compared to the
    predictions of the \emph{maximum a posteriori} model.  The
    \emph{upper panel} shows the galaxy H$\alpha$ rotation curve (blue
    data points): the red line represents the intrinsic model circular
    velocity, while the black line gives the predicted observable,
    i.e. the model circular velocity after the beam-smearing, finite
    slit width and inclination effects are taken into account. The
    light blue data points are not used to constrain the model (see
    text). The \emph{lower panel} shows the model projected second
    velocity moment~$\mu_{2}$ (black line) compared to the
    corresponding observational quantity~$v_{\mathrm{rms}}$ (blue data
    points). The stellar kinematic data sets~$|{v_{\mathrm{rot}}}|$
    (stellar rotation curve, green) and~$\sigma$ (stellar velocity
    dispersion profile, red) are also shown for reference. See
    \Sref{sec:kinematics} for a rigorous definition of these
    quantities.}
  \label{fig:JAMcomp}
\end{figure}
% -------------------------------------------------------------------

\section{Observations}
\label{sec:observations}

In this Section we briefly recall the data set available for this
study. A more detailed description of the data is given in~Paper~II.

 %  IMAGING  %  %  %  %  %  %  %  %  %  %  %  %  %  %  %  %  %  %  % % 

\subsection{Imaging data} 
The imaging data consists of a high spatial resolution (FWHM $\simeq
0.15$ arcsec) K'-band image taken with adaptive optics on the Keck II
telescope. The galaxy-subtracted image (see Paper~II) is used for the
strong lensing analysis, while the light profile of the galaxy is
fitted with a set of elliptical Gaussians which are deconvolved and
deprojected to provide a 3D model of the stellar mass (up to the
normalization), as detailed in \Sref{ssec:mass-model:light}. The
lens-subtracted image used as data set for the lensing analysis is
shown in the upper-right panel of \Fref{fig:LENcomp}. Multi-band HST
photometry is also available and used to determine stellar mass as
discussed in Papers~I and~II.

 %  KINEMATICS  %  %  %  %  %  %  %  %  %  %  %  %  %  %  %  %  %  %  % % 

\subsection{Kinematic data} 
The second set of data that we will use to constrain our mass model is
the rotation and velocity dispersion profiles derived from optical
emission- and absorption-line spectroscopy. A major axis long-slit
spectrum of {\jdisk} was obtained with the DEep Imaging Multi-Object
Spectrograph (DEIMOS) on the Keck~II 10-m telescope.

We used the 1200 line grating (corresponding to a pixel scale of
$0.32$\,$\AA$) with a $1$~arcsec width slit resulting in a spectral
resolution of $\sim 1.9$\,$\AA$.  The wavelength range was
$5200-7800$\,$\AA$, covering several prominent emission and absorption
lines. At the wavelength of Mgb the velocity resolution is
$\sigma_{\rm res}\simeq 41$ \kms, and for H$\alpha$ it is $\sigma_{\rm
  res} \simeq 32$ \kms.  We took three exposures of 1200s in excellent
seeing conditions of $0.60$~arcsec. The spectra were reduced using
routines developed by D.~Kelson \citep{Kelson2003}.

Kinematic parameters were measured from one-dimensional spectra
extracted along the slit with a spatial sampling of $\simeq 0.59$
arcsec (5 DEIMOS pixels), corresponding to one data point per seeing
FWHM.  The rotation and velocity dispersion profile of the stars were
obtained by fitting a region including the Mgb [5177\,$\AA$] and FeII
[5270\,$\AA$] lines with a set of stellar templates.

The rotation curve of the ionized gas was measured by fitting
Gaussians to the H$\alpha$ line [6563\,$\AA$], and is shown in the
upper panel of \Fref{fig:JAMcomp} (data points with error
bars). Outside of the inner $\sim 2$ arcsec the velocity dispersion of
the H$\alpha$ line was equal to the instrumental resolution,
indicating the ionized gas disk is dynamically cold.

In our dynamical model (see \Sref{sec:kinematics}), we assume that the
ionized gas traces the circular velocity of the galaxy (i.e., there is
no pressure support).  For the stellar kinematics our model implicitly
includes rotation and dispersion, although neither of these parameters
are fitted to directly. Instead, our model predicts the projected
second velocity moment, which is fitted to the root mean square
velocity of the stars, $v_{\mathrm{rms}}(R)=
\sqrt{v_{\mathrm{rot}}^2(R) + \sigma^2(R)}$
\citep[see][]{Cappellari2008}. The lower panel of \Fref{fig:JAMcomp}
shows the observables~$v_{\mathrm{rot}}$ and~$\sigma$ (as green and
red data points, respectively) as well as the root mean square
velocity (blue data points).

In our modelling, as described in Paper~II, we conservatively exclude from
the fit the data points of the gas rotation curve that are (i) within
the inner 2~arcsec (due to uncertainties and likely asymmetries in the
H$\alpha$ distribution in this region) and (ii) beyond $3.5$~arcsec on
the west side of the rotation curve (where there is an asymmetry
caused by the presence of the warp). These excluded points are shown
in light blue in \Fref{fig:JAMcomp}.

%%%%%%%%%%%%%%%%%%%%%%%%%%% MASS MODEL %%%%%%%%%%%%%%%%%%%%%%%%%%%%%%%

\section{The galaxy mass model}
\label{sec:mass-model}

In order to perform a self-consistent analysis of the mass structure
of {\jdisk} we need to combine the constraints derived from both the
lensing and kinematics data sets. The most general and straightforward
way to proceed is simply to adopt for the galaxy a plausible total
mass density distribution $\rho_{\mathrm{tot}}(\vec{x},\veceta)$,
where $\vec{x}$ denotes the spatial coordinates and $\veceta$ is a set
of parameters characterizing the density profile, and use it to model
simultaneously the various sets of observables. The main challenge
with this approach lies in choosing a mass distribution that is
realistic and flexible enough to reproduce the data, but at the same
time simple enough that the exploration of the $\veceta$ parameter
space remains computationally feasible. 

In keeping with local studies of both disk and elliptical galaxies
(e.g.~\citealt{Weijmans2008}, \citealt{Williams2009}) we model the
mass distribution using two components: (i) a luminous mass component
whose detailed profile is obtained by deprojecting the observed
surface brightness distribution and (ii) a generalized NFW dark matter
halo whose profile is motivated by cosmological simulations.

Throughout the remainder of this work we will make reference to the
two following right-handed coordinate systems: (i) a cylindrical
coordinate system $(R, \varphi, z)$ with the $z$-axis directed along
the galaxy rotation axis; (ii) a Cartesian coordinate system $(x', y',
z')$, where the $z'$-axis is directed along the line-of-sight and
$(x', y')$ denotes the plane of the sky, with the $x'$-axis aligned
along the galaxy projected major axis. In both cases the origin of the
axes is located in the center of the galaxy (which is assumed to be an
axially symmetric system). We use the first reference system to write
the intrinsic galactic quantities (e.g.\ density and potential) and
the second one to express the projected quantities (e.g.\ surface
brightness and deflection angle).  We denote with~$i$ the galaxy
inclination, i.e. the angle comprised between the $z$- and $z'$-axes
(so that $i = 90^{\circ}$ for a system observed edge-on).

 %  STELLAR MASS  %  %  %  %  %  %  %  %  %  %  %  %  %  %  %  %  %  %

\subsection{Luminous mass distribution}
\label{ssec:mass-model:light}

An ideal model for the luminous mass distribution should be flexible
and realistic (in particular, it should be able to reproduce the
observed surface brightness distributions when projected along the
line-of-sight), and analytically simple, so that the corresponding
gravitational potential is easy to calculate.  This can be achieved by
making use of the MGE method, a technique originally pioneered by
\citet{Bendinelli1991} and subsequently generalized and developed by
\citet*{Monnet1992}, \citet*{Emsellem1994}, \citet*{Emsellem1999} and
\citet{Cappellari2002}, whose formalism we follow here. In order to
minimize dust obscuration and map as closely as possible the stellar
distribution, we apply the MGE decomposition to the high-resolution
K'-band image of the {\jdisk} surface brightness distribution.

The observed galaxy surface brightness $\Sigma (x',y')$ is
parametrized as a sum of~$N$ two-dimensional, concentric,
elliptically-symmetric Gaussian components $g_{k} (x',y')$, each with
luminosity $L_{k}$:
\begin{equation}
  \label{eq:mge:Sigma}
  \Sigma(x',y') = \sum_{k} L_{k} \, g_{k} (x',y') \, ,
\end{equation}
where each Gaussian function
\begin{equation}
  \label{eq:mge:gaussian}
  g_{k} (x',y') = \frac{1}{2 \pi \sigma^{2}_{k} q'_{k}}
                         \exp \left[ - \frac{1}{2 \sigma_{k}^{2}} 
                         \left( {x'}^{2} + \frac{{y'}^{2}}{{q'_{k}}^{2}} 
                         \right) \right]
\end{equation}
is characterized by the widths $\sigma_{k}$ and $q'_{k} \sigma_{k}$
along the $x'$- and $y'$-axis respectively, and $q'_{k}$ is the
projected axial ratio of the $k$-th component. The total stellar
luminosity of the system is simply given by $L_{\mathrm{tot}} =
\sum_{k} L_{k}$.

%\footnote{We note that, since the individual components $g_{k}$
%  do not have a direct physical meaning, $L_{k}$ can be simply
%  interpreted as weights and are, therefore, allowed to be negative,
%  provided that the total luminosity and density are positive.}.

In general, even assuming\,---\,as we do\,---\,that the galaxy
inclination angle~$i$ is known, the deprojection of the observed light
distribution of an axisymmetric galaxy is an intrinsically degenerate
problem unless the system is seen edge-on \citep{Rybicki1987}. The
solution, however, can become unique when a model is specified. In the
case of the MGE parametrization, the deprojected three-dimensional
luminosity density distribution has the simple expression
\begin{equation}
  \label{eq:mge:rho}
  \rho(R,z) = \sum_{k}
              \frac{L_{k}}{(2 \pi)^{3/2} \sigma_{k}^{3} q_{k}}
              \exp \left[ - \frac{1}{2 \sigma_{k}^{2}} \left(
              R^{2} + \frac{z^{2}}{q_{k}^{2}} \right) \right] \, ,
\end{equation}
which is still a sum of Gaussian functions with intrinsic axial ratios
given by
\begin{equation}
  \label{eq:mge:q}
  q^{2}_{k} = \frac{{q'_{k}}^{2} - \cos^{2} i}{\sin^{2} i} \, .
\end{equation}
Since the stellar component of galaxies is oblate or spherical, most
(if not all) Gaussians will turn out to have~$0 \le q_{k} \le 1$. In
order for the axial ratios of the 3D Gaussian components to be
physical, one must enforce the constraint that the projected axial
ratios~$q'_{k}$ are rounder than~$|{\cos i}|$ when fitting the profile
of \Eqref{eq:mge:Sigma} to the observed surface brightness
distribution.

The luminosity density in \Eqref{eq:mge:rho} can be straightforwardly
converted into a mass density by multiplying each term by the stellar
mass-to-light ratio $\MoL_{k}$, so that the mass of each Gaussian is
given by $M_{k} \equiv \MoL_{k} L_{k}$.  However, the single Gaussian
elements are simply a mathematically convenient way to describe the
light profile and do not have a direct physical meaning
individually. Therefore, since there is little interest in studying
them one by one, in this work we assign the same global stellar
mass-to-light ratio $\MoLs$ to all the luminous components. This
simplifying assumption is equivalent to assuming that the bulge and
disk components have the same stellar mass-to-light ratio. Note that
the choice of the K$'$-band image as trace of stellar light should
minimize variations in mass-to-light ratio.  Future work with higher
resolution data should explore further the limitations of this
assumption.

An additional advantage of the MGE approach is that we avoid dealing
with the difficult and somewhat degenerate problem of decomposing the
light profile into the separate disk and bulge contributions
\citep[see e.g.][]{vanderKruit-Searle1981} since we have a model that
can fit very accurately the whole light distribution at once.

The density distribution of \Eqref{eq:mge:rho} is a sum of components
stratified on homoeoidal surfaces, hence the corresponding
gravitational potential can be derived using the classic
\citet{Chandrasekhar1969} formula, obtaining
\citep[see][]{Emsellem1994}
\begin{equation}
  \label{eq:mge:pot}
  \Phi(R,z) = - G \sqrt{\frac{2}{\pi}} \, \sum_{k}
              \frac{M_{k}}{\sigma_{k}} \,
              \tilde{\Phi}_{k}(R,z) \, ,
\end{equation}
where $G$ is the gravitational constant and the dimensionless function
\begin{equation}
  \label{eq:mge:pot_tilde}
  \tilde{\Phi}_{k}(R,z) = \int_{0}^{1} 
        \frac{\de \tau}{\sqrt{1 - \eta_{k}^{2} \tau^{2}}}
        \exp \left[ - \frac{\tau^{2}}{2 \sigma_{k}^{2}} \left(
        R^{2} + \frac{z^{2}}{1 - \eta_{k}^{2} \tau^{2}}
        \right) \right],
\end{equation}
\noindent
with $\eta_{k}^{2} \equiv 1 - q_{k}^{2}$, can be evaluated with a
single numerical integral.  The density distribution
(Eq.~\ref{eq:mge:rho}) and its potential (Eq.~\ref{eq:mge:pot}) are
remarkably simple for such a flexible mass model. Even better, the
corresponding kinematic quantities, obtained by solving the Jeans
equations, also have relatively straightforward expressions that do
not involve any special functions (see \citealt{Cappellari2008} for a
rigorous derivation of the velocity moments).

 %  DM HALO  %  %  %  %  %  %  %  %  %  %  %  %  %  %  %  %  %  %  %

\subsection{Dark matter halo}
\label{ssec:mass-model:DM}

Cold dark matter simulations are known to produce halos with, on
average, universal mass density profiles \citep*[][NFW]{Navarro1997}
that are well fitted by a broken power-law functional form with an
inner logarithmic slope~$\slope = 1$ and a slope~$\slope = 3$ in the
outer regions, i.e. at radial distances much larger than the scale
radius~$\rs$. The situation, however, becomes far more complex when
the baryons are added to the picture and\,---\,although the detailed
mechanisms are not yet fully understood\,---\,it is widely accepted
that the involved processes can have the effect of modifying the inner
slope of the dark halo density profile \citep[e.g.][]{Blumenthal1986,
  Dekel2003, Gnedin2004, Nipoti2004, Abadi2010}. Therefore, in order
to account for a dark matter halo that can be either steeper or
shallower than a NFW in the inner regions, we adopt an axisymmetric
generalized NFW (gNFW) density distribution
\citep[see][]{Zhao1996,Wyithe2001}:
\begin{equation}
  \label{eq:rho_gNFW}
  \rhoDM(m) = 
     \frac{\delta_{c} \, \rho_{\mathrm{crit}}}
          {(m/\rs)^{\slope}\,(1 + m/\rs)^{3-\slope}} \, .
\end{equation}
Here, $\rho_{\mathrm{crit}}$ is the critical density of the universe
at the redshift of the object in question, and $m$ denotes the
elliptical radius, i.e.
\begin{equation}
  \label{eq:m}
  m^2 \equiv R^{2} + \frac{z^2}{{\qh}^{2}} ,
\end{equation}
where $\qh$ indicates the three-dimensional axial ratio of the profile
(the halo is oblate for \mbox{$\qh < 1$}, spherical for \mbox{$\qh =
  1$}, prolate for \mbox{$\qh > 1$}).  Note that we take the halo to
be aligned with the stellar mass distribution, as in Paper~II.

In order to enable an easier comparison of the scale radii between
profiles having different values of~$\slope$, it is useful to
introduce the quantity~$\rsg \equiv (2-\slope)\rs$, which corresponds
to the radius at which the logarithmic density slope of the profile
is~$-2$. Clearly, $\rsg = \rs$ only in the case of the regular NFW
profile. Another useful scale length is the ``virial'' radius~$\rvir$,
defined as the spherical radius within which the average density is
equal to 200~$\rho_{\mathrm{crit}}$. The concentration parameter of
the halo is usually expressed as the ratio $c = \rvir/\rs$; an
alternative definition, adopting the radius~$\rsg$, is $\cg =
\rvir/\rsg$.

The characteristic halo density $\delta_{c}$ that sets the
normalization of~$\rhoDM$ in the center is then 
expressed (following, e.g.,
\citealt{Dutton2005})\footnote{Note that there is a typographical
error in Eq.~(7) of \citet{Dutton2005}: inside the integral the
numerator should read $y^{2-\alpha} [1 + (2-\alpha)y]^{\alpha-3}$.} 
as a function of both the concentration and the slope:
\begin{equation}
  \label{eq:deltac}
  \delta_{c} = \frac{200}{3} \frac{c^{3}}{\zeta(c,\slope,1)} \, ,
\end{equation}
where we have defined the function
\begin{equation}
  \label{eq:zeta}
  \zeta(c,\slope,\qh) = \int_{0}^{\,c} 
  \frac{\tau^{2-\slope} (1+\tau)^{\slope-3}}
       {\sqrt{1 - (1-\qh^{2}) \tau^{2}/c^{2}}} \, \de \tau \, .
\end{equation}

The mass distribution given by \Eqref{eq:rho_gNFW} is completely
specified when the four independent parameters~$\slope$, $\qh$, $\rs$
and~$c$ are given. In this work, we choose to re-parametrize the halo
using the virial velocity $\vvir$, i.e. the circular velocity at the
virial radius, in place of the scale radius, since~$\vvir$ has a very
intuitive physical interpretation and facilitates the comparison with
theoretical work, where this quantity is frequently employed
\citep[see e.g.][]{Maccio2008,Dutton2011-DH}. If the velocity is
expressed in km s$^{-1}$ and the radii in kpc, then one can show
(cf.\ \citealt{Dutton2005}) that $\vvir$ is related to the virial
radius by the formula
\begin{equation}
  \label{eq:vvir}
  \left( \frac{\vvir}{\rvir} \right)^{2} = 
  h^{2} \, \frac{\zeta(c,\slope,\qh)}{\zeta(c, \slope, 1)} \, ,
\end{equation}
where $h = H/100$\kms Mpc$^{-1}$ and $H$ denotes the value of the
Hubble constant at the redshift of the object.

It is convenient to perform an MGE of the axisymmetric gNFW profile in
order to simplify considerably both the calculation of the lensing
angle and the solving of the Jeans equations (cf., e.g.,
\citealt{Williams2009}, where a MGE decomposition of the NFW halo is
performed). We find that around~8 Gaussian components are typically
enough to provide an excellent fit to both the $\rhoDM$ distribution
and the lensing deflection field (typically within 1-3 per cent),
ensuring that the adoption of this approximation does not change our
inferences. In this case, the total potential is still obtained from
\Eqref{eq:mge:pot} by extending the sum to include also the
$N_{\mathrm{DM}}$ Gaussian elements that describe the dark halo
component.

%%%%%%%%%%%%%%%%%%%%% LENS MODELING %%%%%%%%%%%%%%%%%%%%%%%%%%%%%%%%%%

\section{Modelling the gravitational lensing} 
\label{sec:lensing}

Given the observed surface brightness distribution of the lensed
images and a mass model for the deflector, we recover the unlensed
surface brightness distribution of the background object (the
``source'' object) by making use of the pixellated source
reconstruction method, which takes into account the effects of PSF
blurring and regularization \citep[see
  e.g.][]{Warren-Dye2003,Koopmans2005,Suyu2006,Brewer-Lewis2006a}.
Our implementation of this method is described in detail in
\citet{Barnabe-Koopmans2007} and is included in the {\cauldron} code
that has been employed in the combined lensing and dynamics analysis
of the SLACS early-type galaxies for which two-dimensional kinematic
maps are available (see \citealt{Czoske2008, Czoske2012};
\citealt{Barnabe2009}, \citeyear{Barnabe2008},
\citeyear{Barnabe2011}).

This approach consists in casting back, pixel by pixel, through the
lensing equation, the lensed image grid onto the source image
grid. The results of this procedure are encoded in the lensing
operator~$\mat{A}$, which allows one to express the mapping of the
background source~$\vec{s}$ into the lensed image~$\vec{d}$ as a
linear problem, i.e.\ $\mat{A} \vec{s} = \vec{d}$. This set of linear
equations is then solved for~$\vec{s}$ by means of very efficient
standard techniques.

All that is needed to calculate the lensing operator is the deflection
angle~$\vec{\alpha}$, which is obtained from the surface mass density
distribution of the lens galaxy \citep*[see e.g.][]{SaasFee2006}.
Therefore, the matrix~$\mat{A}$ depends both on the physical
parameters~$\veceta$ characterizing the density profile and on the
geometry of the system, i.e.\ the inclination~$i$ and the angular
diameter distances between the observer and the source ($\Ds$),
between the observer and the lens ($\Dd$), and between the lens and
the source ($\Dds$).

For many three-dimensional density profiles of astrophysical interest
the deflection angle is very cumbersome to compute (cf., e.g., the
catalogue of \citealt{Keeton2001C}). This has contributed to the
widespread adoption of those few profiles, such as the isothermal
ellipsoid, for which analytical expressions of~$\vec{\alpha}$ are
available \citep*[see][]{Kormann1994,Keeton-Kochanek1998}. Remarkably,
the lensing deflection angle corresponding to the density distribution
of \Eqref{eq:mge:rho} is very straightforward to calculate, involving a
single quadrature and no special functions:
\begin{eqnarray}
  \label{eq:alpha:x}
  \alpha_{x'} (x', y') & = & \frac{1}{\pi \Dd^{2} \Scr} 
                \int_{0}^{1} \tau \, \de \tau 
                \sum_{k} \frac{M_{k}}{\sigma_{k}} 
                \frac{\xt'}
                     {\sqrt{1 - \eta_k^2 \tau^{2}}}
                \nonumber \\
                & & \times \exp \left[ - \frac{\tau^{2}}{2} \left( 
                \xt'^{2} + \frac{\yt'^{2}}{1 - \eta_k^2 \tau^{2}}
                \right) \right] \, ,
\end{eqnarray}
\begin{eqnarray}
  \label{eq:alpha:y}
  \alpha_{y'} (x', y') & = & \frac{1}{\pi \Dd^{2} \Scr} 
                \int_{0}^{1} \tau \, \de \tau 
                \sum_{k} \frac{M_{k}}{\sigma_{k}} 
                \frac{\yt'}
                     {{\left(1 - \eta_k^2 \tau^{2}\right)}^{3/2}}
                \nonumber \\
                & & \times \exp \left[ - \frac{\tau^{2}}{2} \left( 
                \xt'^{2} + \frac{\yt'^{2}}{1 - \eta_k^2 \tau^{2}}
                \right) \right] \, ,
\end{eqnarray}
where both the deflection angle and the widths~$\sigma_{k}$ are
expressed in radians; $\xt' \equiv x'/\sigma_{k}$ and $\yt' \equiv
y'/\sigma_{k}$ are the sky coordinate (normalized to~$\sigma_{k}$)
with respect to the lens center. As before, $\eta_k^2 = (1-q_k^2)$. 
The critical surface density
\begin{equation}
  \label{eq:Scr}
  \Scr = \frac{c^{2}}{4 \pi G} \frac{\Ds}{\Dds \Dd}
\end{equation}
is the characteristic surface density used in gravitational lensing
\citep[e.g.,][]{Treu2010review}.

%%%%%%%%%%%%%%%%%%%%% DYNAMICS MODELING %%%%%%%%%%%%%%%%%%%%%%%%%%%%%%

\section{Modelling the kinematics}
\label{sec:kinematics}

 %  VELOCITY DISPERSION  %  %  %  %  %  %  %  %  %  %  %  %  %  %  % 

\subsection{Predicting the observed second velocity moments 
with the anisotropic Jeans equations}
\label{ssec:jeans}

Let us consider a steady-state axially symmetric stellar system
characterized by a distribution function (DF) $f(\vec{x}, \vec{v})$,
where the positions $\vec{x}$ and the velocities $\vec{v}$ are the
phase-space coordinates, and subject to the influence of a total
gravitational potential $\Phit (R,z)$. While the typical observational
data sets do not allow one, in general, to recover the full
six-dimensional DF, it is possible to gain valuable information on the
global dynamical structure of the system by noting that its velocity
moments must satisfy the two Jeans equations \citep{BT08}
\begin{eqnarray}
  \frac{\partial (\rho \bvzq)}{\partial z} + 
  \frac{1}{R} \frac{\partial (R \rho \bvRvz)}{\partial R} & = &
  -\rho \frac{\partial \Phit}{\partial z} 
  \label{eq:jeans:cyl:z}\\
  \frac{\partial (\rho \bvRq)}{\partial R} + 
  \frac{\partial (\rho \bvRvz)}{\partial z} & = &
  - \rho \frac{\partial \Phit}{\partial R} +
  \rho \frac{\bvfq - \bvRq}{R} \, .
  \label{eq:jeans:cyl:R}
\end{eqnarray}

Here, $\rho (R,z) \equiv \int f \de^{3} \vec{v}$ denotes the
three-dimensional density distribution of the stellar system, and the
bar indicates a phase-space average of the quantity of interest, i.e.
\begin{equation}
\overline{v_{i} v_{j}} \equiv \frac{1}{\rho} \int v_{i} v_{j} f \de^{3}
  \vec{v}.
\end{equation}

The system is not required to be self-gravitating and therefore in
Eqs~(\ref{eq:jeans:cyl:z}) and (\ref{eq:jeans:cyl:R}) $\rho$ might
well be the density distribution $\rho_{\mathrm{tr}}$ of a tracer
stellar component described by a DF $f_{\mathrm{tr}}$ and subject to
an external potential.  Moreover, if other collisionless components
(each one defined by its own DF) are present, each one will obey its
own set of Jeans equations \emph{within the same total potential}
$\Phit$.

In our dynamical analysis of late-type galaxies, we adopt an
axisymmetric total potential $\Phit = \Phistar + \PhiDM$, where the
two components represent the potentials of the luminous distribution
and dark matter halo, respectively. We then write down and solve the
Jeans equations, using the stellar density distribution $\rhostar$
associated to the corresponding potential via the Poisson equation, in
order to obtain the intrinsic velocity moments.\footnote{Because of
  the collisionless nature of dark matter, one could write an
  analogous set of equations also for the halo component. However,
  since the corresponding velocity moments cannot be observed, this
  would be of no use in the present context of comparing the model
  predictions with the observed data sets.} These are then projected
along the line-of-sight and\,---\,after taking into account the effect
of instrumental PSF and aperture integration\,---\,compared with the
corresponding observational quantities.

Of course, even when the potential and the density distributions are
given, the two Equations~(\ref{eq:jeans:cyl:z}) and
(\ref{eq:jeans:cyl:R}) still depend on the four unknown functions
$\bvRq$, $\bvfq$, $\bvzq$ and $\bvRvz$, and therefore additional
assumptions are needed in order to determine a unique solution for the
Jeans equations. This is usually achieved by assigning the orientation
and the shape of the intersection of the velocity dispersion ellipsoid
with the meridional plane $(R,z)$ at each point.

Observations of the Milky Way and of nearby disk galaxies show that
the velocity dispersion ellipsoid is more flattened in the vertical
direction than in the radial one\footnote{If the assumption of a
  steady-state axisymmetric system holds, this implies that the disk
  DF also depends on a third, non-classical, integral of motion
  $I_{3}$, in addition to the two classical integrals, namely the
  energy~$E$ and the angular momentum~$\Jz$ along the rotation axis.},
i.e. $\bvzq < \bvRq$ (see e.g.\ \citealt{Wielen1977},
\citealt*{Gerssen1997, Gerssen2000},
\citealt*{vanderKruit-deGrijs1999}, \citealt*{Shapiro2003},
\citealt{Ciardullo2004}, \citealt*{Noordermeer2008}).

\citet[][]{Cappellari2008} introduced a simple and effective way
(referred to as anisotropic Jeans models) to provide a closure for the
Jeans equations that manages to reproduce this important feature. The
two assumptions of this phenomenological model are: (i) the velocity
dispersion ellipsoid is aligned with the cylindrical coordinate system
(so that the mixed terms $\bvRvz$ are everywhere zero) and (ii) the
anisotropy in the meridional plane is constant, i.e. $\bvRq = b
\bvzq$, with the anisotropy parameter $b \ge 0$. The meridional plane
anisotropy is usually expressed in the literature using the
parameter~$\beta_{z}$, such that
\begin{equation}
  \label{eq:beta}
  \beta_{z} = 1 - \frac{\bvzq}{\bvRq} = 1 - \frac{1}{b} \, .
\end{equation}

In real galaxies, the shape and the orientation of the velocity
dispersion ellipsoid are in general a non-trivial function of the
position on the meridional plane. However, the assumption of
cylindrical alignment is quite accurate for fast-rotating galaxies and
disk systems in general, in particular along the minor axis and, more
crucially, in the vicinity of the equatorial plane, where the density
is highest. In fact, Jeans models constructed with this simple
prescription for the anisotropy have been shown to reproduce
remarkably well the observed kinematic moments of fast rotators and
spirals (\citealt{Scott2009}, \citealt*{Williams2009}).

With these assumptions, the Jeans Equations~(\ref{eq:jeans:cyl:z})
and~(\ref{eq:jeans:cyl:R}) become
\begin{eqnarray}
  \frac{\partial (\rho \bvzq)}{\partial z} & = &
  -\rho \frac{\partial \Phit}{\partial z} 
  \label{eq:jeans:aniso:z}\\
  b \frac{\partial (\rho \bvzq)}{\partial R} & = &
  - \rho \frac{\partial \Phit}{\partial R} +
  \rho \frac{\bvfq - b \bvzq}{R} \, .
  \label{eq:jeans:aniso:R}
\end{eqnarray}

From the equations above, and 
imposing the intuitive constraint that the
vertical pressure $\rho \bvzq = 0$ for $z \to \infty$, one obtains 
the following
expressions for the intrinsic second velocity moments along the
coordinate directions:
\begin{eqnarray}
\bvzq & = & \frac{1}{\rho} \int_{z}^{\infty} \rho 
            \frac{\partial \Phit}{\partial z'} \de z'
      \label{eq:vzq}\\
\bvfq & = & \frac{b}{\rho} R \frac{\partial \rho \bvzq}{\partial R} + 
            b \bvzq + R \frac{\partial \Phit}{\partial R} \, .
      \label{eq:vfq}
\end{eqnarray}
These intrinsic quantities are then integrated along the line of sight
to obtain the projected second velocity moment
$\overline{v^{2}_{\mathrm{los}}}$ (whose square root is usually
indicated as~$\mu_{2}$) which can be directly compared to the stellar
kinematics observables. The projected velocity moments for the case of
MGE parametrization are calculated in \citet{Cappellari2008}.  The
observational counterpart of the model quantity $\mu_{2}$ is the root
mean square velocity $v_{\mathrm{rms}} \equiv
\sqrt{v_{\mathrm{rot}}^{2} + \sigma^{2}}$, where $v_{\mathrm{rot}}$
and $\sigma$ indicate the line of sight projected stellar rotation
velocity and velocity dispersion, respectively.

We recall that, given a potential and a density distribution, the
Jeans equations impose a condition for the equilibrium on the second
velocity moments, but they provide no prescription on how to separate
these moments into the contributions of random and ordered
motions. Since no net radial or vertical motions are considered, and
thus $\sigma^{2}_{R} = \bvRq$ and $\sigma^{2}_{z} = \bvzq$, here this
issue would only be relevant for the splitting of the azimuthal
component into the streaming motion~$\bvf$ and the velocity
dispersion~$\sigma_{\varphi}$, i.e. $\bvfq = {\bvf}^{2} + \sigfq $,
which is usually tackled by adopting \emph{ad hoc} assumptions such as
the \citet{Satoh1980} decomposition. In this work, however, we avoid
making any additional assumptions in order to model~$\bvf$ separately,
and we only model the second velocity moments as described above.

 %  ROTATION CURVE  %  %  %  %  %  %  %  %  %  %  %  %  %  %  %  %  % 

\subsection{Predicting the observed gas circular velocity}
\label{ssec:vcirc}

In order to model the rotation curve of the H$\alpha$ gas, we
calculate the circular velocity profile, $\vc(R)$, of a test particle
of negligible mass in a circular orbit in the equatorial plane of the
galaxy.

The circular velocity, as it is clear from its definition,
i.e. $\vc^{2}(R) = R \, (\partial \Phit / \partial R) |_{z=0}$, is
uniquely determined by the total gravitational potential of the
galaxy. In general, $\vc$ differs from the stellar rotation velocity
$\bvf$ (often referred to as streaming motion) which is usually lower
due to the effect of the stellar velocity dispersion, which acts as a
pressure term \citep[see e.g.][]{BT08}.

The (squared) circular velocity profile that corresponds to the MGE
mass model described in \Sref{sec:mass-model} is readily calculated
from Eqs~(\ref{eq:mge:pot})--(\ref{eq:mge:pot_tilde}) and has the
following expression:
\begin{eqnarray}
  \label{eq:vcirc}
\vc^{2} (R) = 4 \pi G R^{2} \int_{0}^{1} \tau^{2} \de \tau \, 
              \sum_{k} \, \frac{q_{k} \, \rho_{0k}}
                               {\sqrt{1 - \eta_k^2 \tau^{2}}}
              \exp \left( - \frac{\tau^{2}}{2 \sigma_{k}^{2}} R^{2} \right) 
              \, ,
\end{eqnarray}
where again $\eta_k^2 = (1-q_k^2)$ 
and the constant $\rho_{0k} \equiv \rho_{k}(0,0)$ is the central
value of the mass density distribution of the $k$-th Gaussian element.

In order to compare the predicted rotation curve~$\vc(R)$ of
\Eqref{eq:vcirc} with the observations, we also take into account the
combined effects of inclination, PSF blurring and finite slit width,
collectively referred to as beam-smearing. Since the exact
distribution of the H$\alpha$ gaseous component is not known from the
observations, we approximate it using the available $K'$-band light
profile instead, which is more accurate than using an exponential disk
model. Additionally, in keeping with Paper~II, we have cautionarily
excluded from the fit two regions of the rotation curve: (i) the inner
2~arcsec, a region where the H$\alpha$ distribution is probably
asymmetric, due to the effect of extinction and (ii) the outermost
three points of the west side of the rotation curve, which are
affected by a spurious decrease of the velocity caused by the warp.

% --------------------- CORNERPLOT: LEN -----------------------------
\begin{figure*}
  \centering
  \resizebox{1.00\hsize}{!}{\includegraphics[angle=0]
            {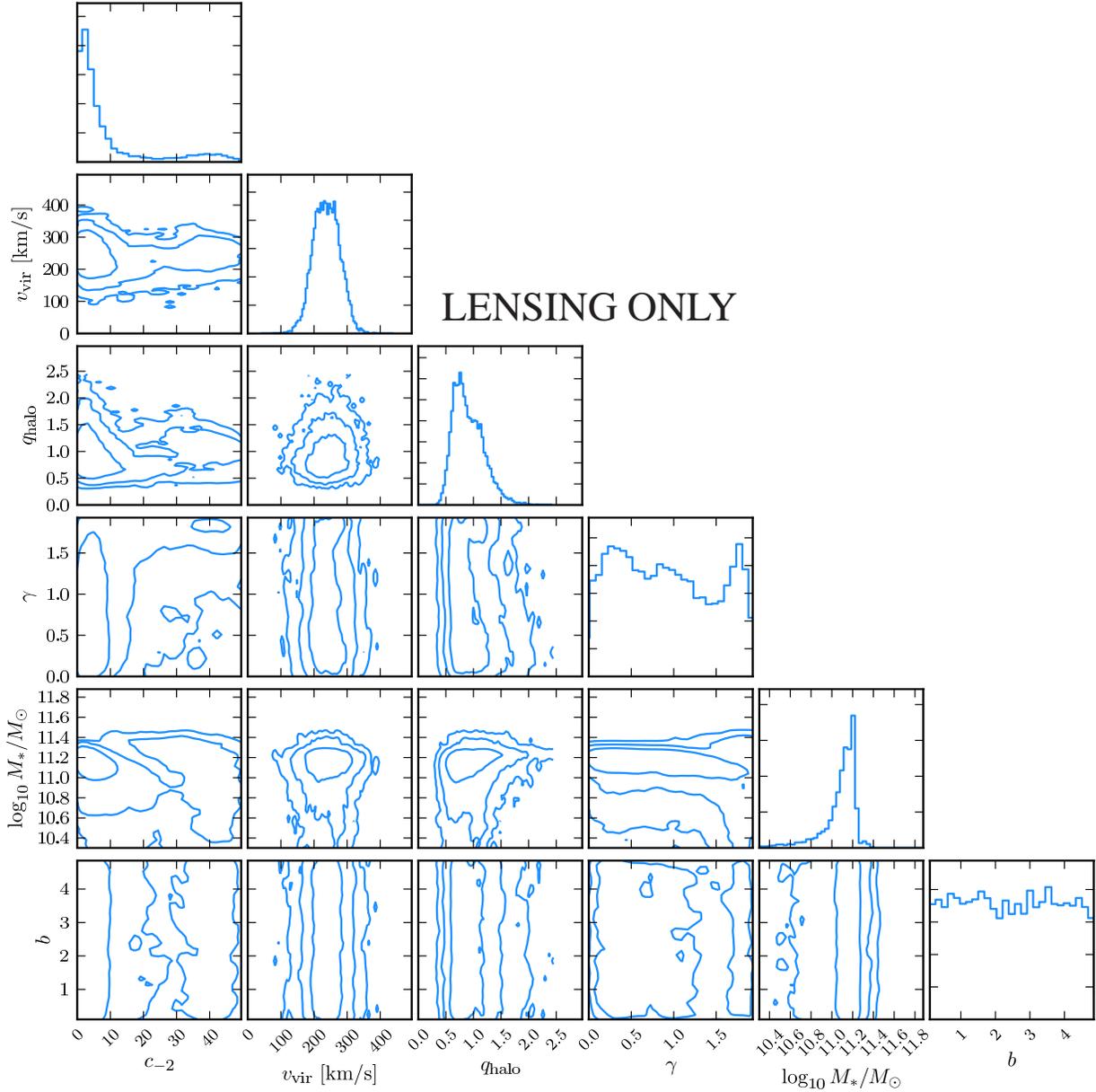}}
  \put(-300,345){\huge LENSING ONLY}
  \caption{Marginalized two-dimensional (contour plots) and
    one-dimensional (histograms) posterior PDFs for the galaxy model
    parameters using constraints from the gravitational lensing data
    set only. The three contours indicate the regions containing,
    respectively, 68\%, 95\% and 99.7\% of the probability.}
  \label{fig:cplot:len}
\end{figure*}
% -------------------------------------------------------------------

% --------------------- CORNERPLOT: DYN -----------------------------
\begin{figure*}
  \centering
  \resizebox{1.00\hsize}{!}{\includegraphics[angle=0]
            {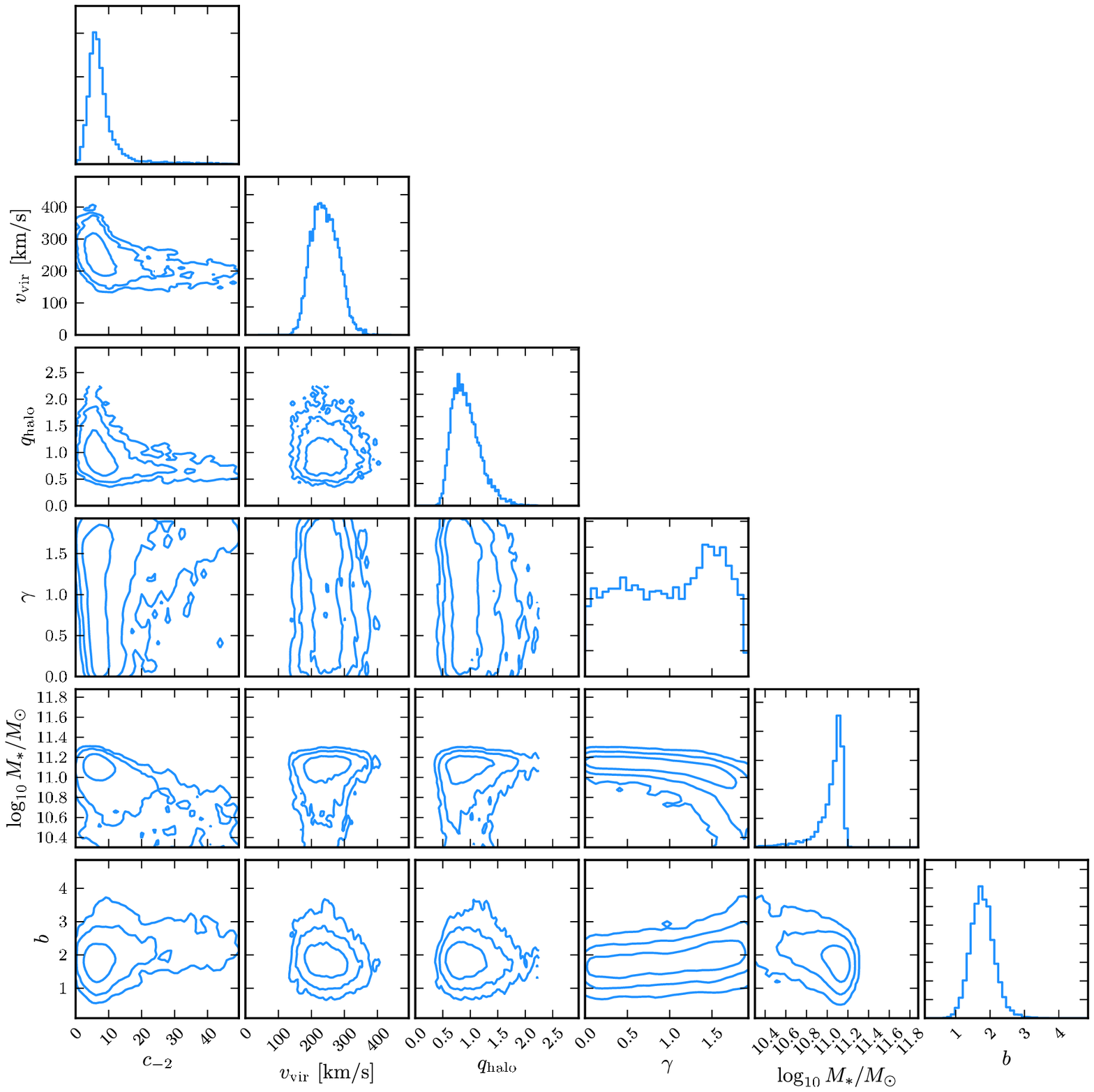}}
  \put(-300,345){\huge KINEMATICS ONLY}
  \caption{Marginalized two-dimensional (contour plots) and
    one-dimensional (histograms) posterior PDFs for the galaxy model
    parameters using constraints from the kinematics data set
    only. The three contours indicate the regions containing,
    respectively, 68\%, 95\% and 99.7\% of the probability.}
  \label{fig:cplot:dyn}
\end{figure*}
% -------------------------------------------------------------------

% --------------------- CORNERPLOT: TOT -----------------------------
\begin{figure*}
  \centering
  \resizebox{1.00\hsize}{!}{\includegraphics[angle=0]
            {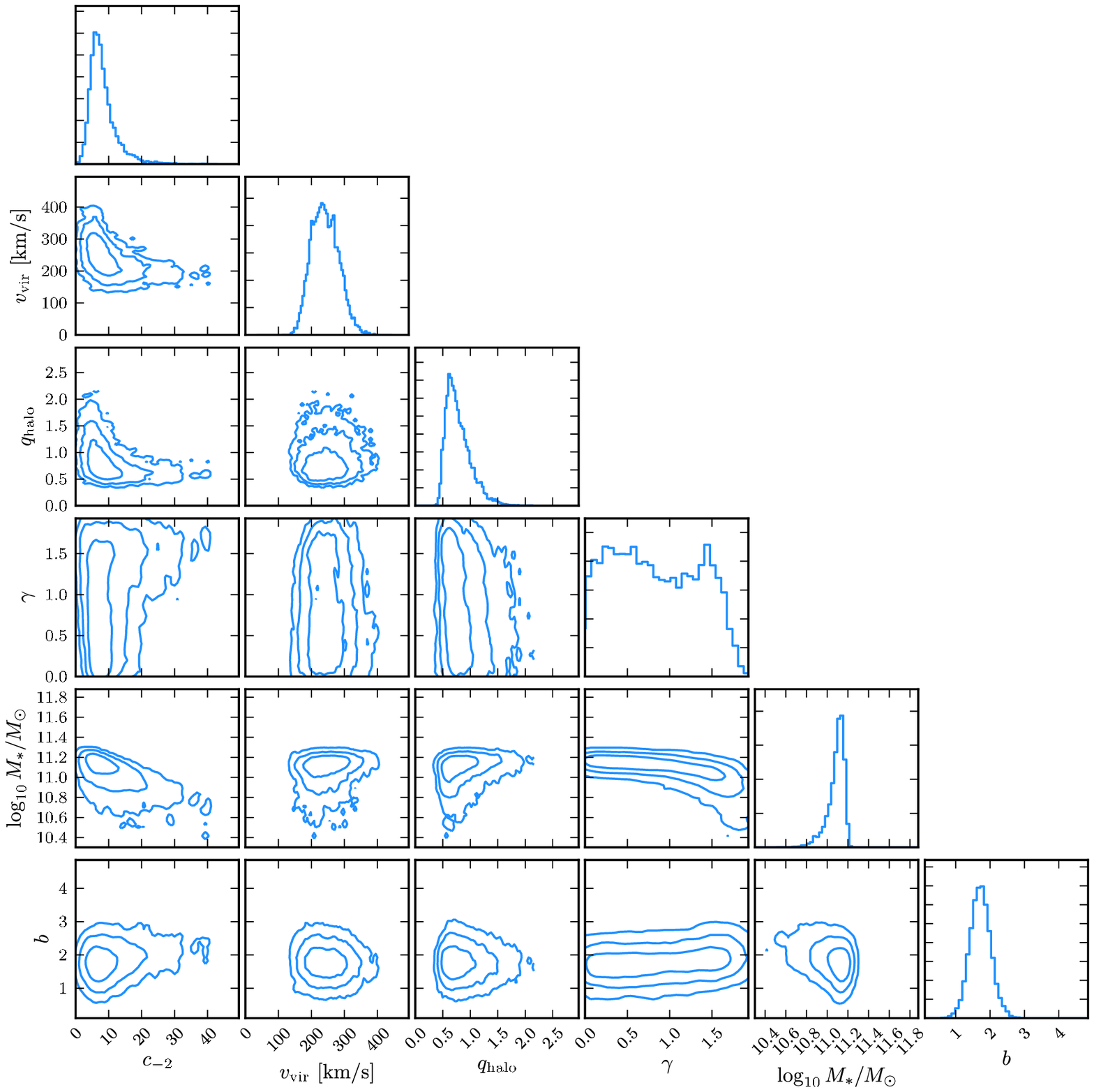}}
  \put(-300,345){\huge LENSING + KINEMATICS}
  \caption{Marginalized two-dimensional (contour plots) and
    one-dimensional (histograms) posterior PDFs for the galaxy model
    parameters using the combined constraints from both lensing and
    kinematics. The three contours indicate the regions containing,
    respectively, 68\%, 95\% and 99.7\% of the probability.}
  \label{fig:cplot:tot}
\end{figure*}
% -------------------------------------------------------------------

%%%%%%%%%%%%%%%%%%%%%%%%%%% INFERENCE %%%%%%%%%%%%%%%%%%%%%%%%%%%%%%%%

\section{Bayesian inference and uncertainties}
\label{sec:bayes}

In order to derive rigorous constraints on the parameters that
characterize the adopted model, we conduct our analysis within the
standard framework of Bayesian statistics (see, e.g.,
\citealt{MacKay2003} and \citealt{SS2006} for an extensive treatment
of this subject).

Let us denote the combined data sets as $\vec{d}$ and the considered
hypothesis as $\mathcal{H}(\vec{\theta})$. In our case, for instance,
$\mathcal{H}$ includes the model that we have adopted to describe the
mass distribution and dynamics of the galaxy under study
(Sections~\ref{sec:mass-model}--\ref{sec:kinematics}), and also all
the assumptions we make about the uncertainties on the data,
instrument response functions and any prior knowledge of the situation
we might want to include.  The non-linear parameters~$\vec{\theta}$
may include, in general, not only the physical parameters $\vec{\eta}$
defining the total mass density distribution, but also the parameters
that characterize the dynamics (e.g. anisotropy) and the geometry
(e.g. inclination) of the system.

From Bayes' theorem, the posterior probability distribution function
(PDF) for the set of parameters~$\vec{\theta}$ is given by
\begin{equation}
  \label{eq:posterior}
  \pr(\vec{\theta} \, | \, \vec{d}, \mathcal{H}) =
  \frac{
  \pr(\vec{d}   \, | \, \vec{\theta}, \mathcal{H}) \, 
  \pr(\vec{\theta} \, | \, \mathcal{H})
  }
  {\pr(\vec{d} \,  | \, \mathcal{H})} ,
\end{equation}
where $\pr(\vec{d} \, | \, \vec{\theta}, \mathcal{H})$ is the
likelihood, $\pr (\vec{\theta} \, | \, \mathcal{H})$ is the prior, and
$\pr(\vec{d} \, | \, \mathcal{H})$, i.e. the factor required to
normalize the posterior over~$\vec{\theta}$, is known as the Bayesian
evidence, which is used in comparing different model forms. When
modelling the lensing and kinematic data we do not keep track of the
value of the evidence, but do make use of it in \Sref{ssec:IMF} below.

The set of parameters~$\vec{\theta}_{\mathrm{MAP}}$ for which the
posterior is maximized identifies the \emph{maximum a posteriori}
(MAP) model. The MAP model can be interpreted as a ``best model'' of
sorts, in the sense that it represents the combination of parameters
that is found to best reproduce the data given our assumptions.  We
adopt it as our reference model for the times when we need to show our
best estimates of the predicted observables (lensed image, rotation
curve, velocity moments) and the reconstructed background lensed
source.

The primary quantities of interest are the marginalized posterior PDFs
for individual parameters $\theta_{i}$ obtained by integrating the
joint posterior PDF over all the other parameters.  These integrals
can be performed most readily if we characterise the joint posterior
by a set of sample parameter values drawn from it. The marginalised
distributions are then readily approximated by histograms of these
samples. When a more compact representation is required, we quote
parameter constraints as the median values of these one-dimensional
histograms, $\vec{\theta}_{\mathrm{med}}$, and quantify our
uncertainty with their 68\% credible intervals (CIs, calculated by
taking the 16th and 84th percentiles).

% ------------------- TABLE OF PRIORS AND RESULTS -------------------
\begin{table*}
  \centering
  \begin{minipage}{0.55\linewidth}
    \caption{Summary of the adopted priors and of the posteriors
      inferred from the combined analysis for the model parameters.}
    \label{table:par} %%% IMPORTANT NOTE: label must be under the caption to give the proper table number!
    \begin{tabular}{lll@{\qquad}l}
      \hline
      parameter              & description                          & prior                   & posterior                \\
      \hline
      $\vvir/\kms$           & dark halo virial velocity            & $\mathcal{N}$(255, 45)  & $ 242_{-39}^{+44} $      \\    
      \noalign{\smallskip}
      $\slope$               & dark halo inner logarithmic slope    & U(0, 2)                 & $ 0.82^{+0.65}_{-0.54} $ \\
      \noalign{\smallskip}
      $\cg$                  & dark halo concentration              & U(0, 50)                & $ 7.7_{-2.5}^{+4.2} $    \\
      \noalign{\smallskip}
      $\qh$                  & dark halo 3D axial ratio             & L$\mathcal{N}$(1, 0.3)  & $ 0.75^{+0.27}_{-0.16} $ \\
      \noalign{\smallskip}
      $\Ms/10^{11} \Msun$    & stellar mass                         & U(0, 5)                 & $ 1.32^{+0.16}_{-0.25} $ \\
      \noalign{\smallskip}
      $b$                    & orbital anisotropy parameter: $\orb$ & U(0, 5)                 & $ 1.77_{-0.29}^{+0.30} $ \\
      \hline
    \end{tabular}
    \textit{Note.} In the prior column, $U(a,b)$ denotes a uniform
    distribution over the open interval $(a, b)$; $\mathcal{N}(a,b)$
    denotes a normal distribution, with $a$ being the central value and
    $b$ being the standard deviation; $L\mathcal{N}(a,b)$ denotes a
    lognormal distribution, with $a$ being the central value for the
    variable, and $b$ being the standard deviation for the log of the
    variable. In the posterior column, we list, for each parameter, the
    median value of the corresponding marginalized posterior PDF and the
    uncertainty quantified by taking the 68\% credible interval (i.e.,
    the 16th and 84th percentiles).    
  \end{minipage}
\end{table*}
% -------------------------------------------------------------------

The model that we employ for the analysis of {\jdisk} has six free
parameters, i.e. parameters with uninformative priors which are
allowed to vary and for which the posterior exploration is
performed. These are: the virial velocity~$\vvir$, the inner
logarithmic slope~$\slope$, the concentration~$\cg$ and the
three-dimensional axial ratio~$\qh$ which describe the gNFW dark
matter halo (\Sref{ssec:mass-model:DM}); the global stellar
mass-to-light ratio~$\MoLs$ of the luminous component
(\Sref{ssec:mass-model:light}), that is more readily interpreted when
expressed in terms of the total stellar mass~$\Ms \equiv \MoLs
L_{\mathrm{tot}}$; and the meridional plane anisotropy parameter~$b$
(\Sref{ssec:jeans}). In analogy with Paper~II, we adopt a broad
Gaussian prior for~$\vvir$ centered on 255~km~s$^{-1}$ with a width of
45~km~s$^{-1}$. This corresponds to the prior adopted in Paper~II for
the virial velocity of their non-singular isothermal halo, and is
equivalent to assuming that the scale radius is not so large that the
virial velocity is dramatically larger than the observed rotation
velocity. We also adopt, as in Paper~II, a lognormal prior centered on
$\qh = 1$ (spherical) for the axial ratio, which allows for both
oblate and prolate haloes. We let~$\cg$ vary on a wide uniform prior
from~0 to~50 to represent our ignorance of the halo concentration. The
inner logarithmic slope $\slope$ is allowed to vary between~0 (flat
core) and~2 (isothermal). We note that $\vvir$ is allowed to go to
zero, so that we are also including in our analysis the case of a disk
galaxy with no dark matter halo, fully described by a self-gravitating
stellar mass distribution. By letting the total stellar mass vary
(with uniform prior) between~0 and~$5 \times 10^{11} M_{\sun}$ we
allow for a wide range of contributions of the luminous components to
the total mass, including the limiting case in which the galaxy is
fully dark matter dominated and the stars are only a tracer with
negligible mass. Finally, the anisotropy parameter can vary uniformly
from $b = 0$, indicating a velocity dispersion ellipsoid without any
radial component, to $b = 5$, for which the velocity dispersion
ellipsoid is very elongated along the radial direction: this interval
is wide enough to include all the values of meridional plane
anisotropy observed in real disk galaxies (see \Sref{ssec:beta}). The
model parameters, together with the adopted priors, are summarized in
\Tref{table:par}. A full description of the model also includes a
number of additional parameters that do not represent physical
characteristics of the system (i.e., the line-of-sight inclination,
the lens center, the regularization level, the weights, widths and
flattenings of the individual MGE Gaussians): these are treated as
nuisance parameters and kept fixed or marginalized over.

For the likelihood function we follow the standard approach of
assuming Gaussian errors on the data points \citep[see
  e.g.][]{Brewer-Lewis2006a, Suyu2006, Marshall2007}. In this case,
the joint likelihood can be written simply as
\begin{eqnarray}
  \label{eq:likelihood}
  \lefteqn{
  \pr(\vec{d}   \,  |  \, \vec{\theta}, \mathcal{H}) \propto
  \exp \left\{ 
   -\frac{1}{2} \sum_{i=1}^{N_{\ell}}
    \frac{{\left[ \ell_{i}^{\,\mathrm{obs}} - \ell_{i}^{\,\mathrm{mod}}
                (\vec{\theta}) \right]}^{2}}
         {\sigma_{\ell,\,i}^{2}} \right.
  }
  \nonumber \\
  & & \left.
  -\frac{1}{2} \sum_{i=1}^{N_{\mu_{2}}}
   \frac{{\left[ \mu_{2,\,i}^{\,\mathrm{obs}} - \mu_{2,\,i}^{\,\mathrm{mod}}
               (\vec{\theta}) \right]}^{2}}
        {\sigma_{\mu_{2},\,i}^{2}}
  -\frac{1}{2} \sum_{i=1}^{N_{\vc}}
   \frac{{\left[ v_{c,\,i}^{\,\mathrm{obs}} - v_{c,\,i}^{\,\mathrm{mod}}
               (\vec{\theta}) \right]}^{2}}
        {\sigma_{\vc,\,i}^{2}} \right\} \, ,
\end{eqnarray}
where the three terms inside the exponential represent the familiar
$\chi^{2}$ misfit functions for the separate contributions of
gravitational lensing, stellar kinematics and gas kinematics,
respectively. We indicate as $\ell_{i}^{\,\mathrm{obs}}$ the
$N_{\ell}$ data points constituting the lensing data set, i.e. the
pixel values in the galaxy-subtracted observed lensed image (see
top-right panel of \Fref{fig:LENcomp}), each characterized by an
uncertainty $\sigma_{\ell,\,i}$. We denote as
$\ell_{i}^{\,\mathrm{mod}}$ the corresponding pixel values of the
model-predicted image, which are determined by the specific choice of
model parameters~$\vec{\theta}$ (for example, the bottom left panel of
\Fref{fig:LENcomp} shows $\ell_{i}^{\,\mathrm{mod}}
(\vec{\theta}_{\mathrm{MAP}})$, i.e. the model-predicted image in the
case of the MAP model). An analogous notation (i.e., observed values,
model-predicted values, uncertainties on the data points) applies for
the velocity moment~$\mu_{2}$ in the case of the stellar kinematics
and for the circular velocity~$\vc$ in the case of gas kinematics.

The computationally expensive task of exploring and sampling the joint
posterior distribution is accomplished by making use of the very
efficient and robust \textsc{MultiNest} algorithm
\citep{Feroz-Hobson2008, Feroz2009}, which implements the nested
sampling Monte Carlo technique \citep{Skilling2004, SS2006}, and can
provide reliable posterior inferences even in presence of multi-modal
and degenerate multivariate distributions. For the analysis of
{\jdisk}, we have launched \textsc{MultiNest} with~2000 live points
(the live or active points are the initial samples, drawn from the
prior distribution, from which the posterior exploration is started).
The large number of live points adopted for this study (cf., e.g., the
\textsc{MultiNest} analysis in \citealt{Barnabe2011}) has allowed us
not only to produce very detailed posterior distributions, but also to
gauge the minimum number of live points (which is found to be
$\sim$200) needed to obtain reliable posterior PDFs, which will be
very useful in reducing the computational load in future analyses of
further SWELLS systems.

%%%%%%%%%%%%%%%%%%%%%%%%%%% RESULTS %%%%%%%%%%%%%%%%%%%%%%%%%%%%%%%%%%

\section{Results and discussion}
\label{sec:results}

In this Section we present and discuss the results of our analysis of
the disk galaxy {\jdisk}, combining the constraints from
both the gravitational lensing and the kinematic data sets as
described in the previous Sections.

 %  Galaxy model pars  %  %  %  %  %  %  %  %  %  %  %  %  %  %  %  % 

\subsection{Inferences on the galaxy model parameters}
\label{ssec:inferences}

As discussed in \Sref{sec:bayes}, the inferences on the model
parameters obtained from our analysis are expressed in the form of a
multivariate posterior PDF. We consider six free parameters: the
virial velocity~$\vvir$, inner logarithmic slope~$\slope$,
concentration~$\cg$ and three-dimensional axial ratio~$\qh$ of the
gNFW dark matter halo, the total stellar mass~$\Ms$, and the
meridional plane orbital anisotropy ratio~$b$. Since visualizing the
full six-dimensional surface is challenging, we present the inferences
in the familiar form of ``cornerplots'' that show all possible
one-dimensional and two-dimensional marginalized posterior PDFs for
the six model parameters. The inferences obtained when using just one
single data set are presented in Figures~\ref{fig:cplot:len}
(gravitational lensing only) and~\ref{fig:cplot:dyn} (kinematics
only), while \Fref{fig:cplot:tot} shows the results for the combined
lensing and kinematics data sets. In each plot, the three contours
indicate the regions containing, respectively, 68\%, 95\% and 99.7\%
of the probability, i.e. they represent the analogue of the 1, 2 and
3$\sigma$ contours of a Gaussian distribution. The median value and
the corresponding uncertainties (expressed as 16th and 84th
percentiles) for each individual parameter are listed in
\Tref{table:par}.

The constraints provided by kinematics alone are in general slightly
better than the constraints obtained with a pure gravitational lensing
analysis, in particular for the concentration and the stellar mass;
obviously, the anisotropy parameter~$b$ is completely unconstrained in
the lensing analysis, and thus, in this case, the posterior is nothing
else than the input uniform prior. The inferences on the remaining
parameters have uncertainties of similar magnitude in the two cases,
but the marginalized posterior PDFs have different shapes (note, in
particular, the profile for the marginalized PDF of the halo axial
ratio in the two cases), which makes it possible to tighten the
inferences when lensing and kinematics data are considered
simultaneously. The effectiveness of the combined analysis can be seen
in \Fref{fig:cplot:tot}: in particular, we can place tight
constraints on~$\Ms$ by clipping both the low-mass and the high-mass
tails, and we also improve significantly our inferences on the dark
halo parameters~$\qh$ and~$\slope$, for which we weed out the higher
values. The meaning and implications of the constraints on the model
parameters are discussed below in
Sections~\ref{ssec:fDM}--\ref{ssec:beta}.

The high-probability models drawn from the posterior PDF of the
combined analysis, and in particular the MAP model, reproduce both the
lensing and the kinematic observables very accurately
(Figures~\ref{fig:LENcomp} and~\ref{fig:JAMcomp},
respectively). Similarly to what was found in Paper~II, the most probable
lensing models predict a faint counterimage whose presence is
consistent with the noise level. For the kinematics, we note that the
predicted gas rotation curve matches quite well also the data points
within the inner 2~arcsec, which were conservatively excluded from the
fit.

 %  Galaxy model pars  %  %  %  %  %  %  %  %  %  %  %  %  %  %  %  % 

% ---------------- COMBINED CIRCULAR VELOCITY PROFILE ---------------
\begin{figure*}
  \centering
  \resizebox{0.95\hsize}{!}{\includegraphics[angle=-90]
            {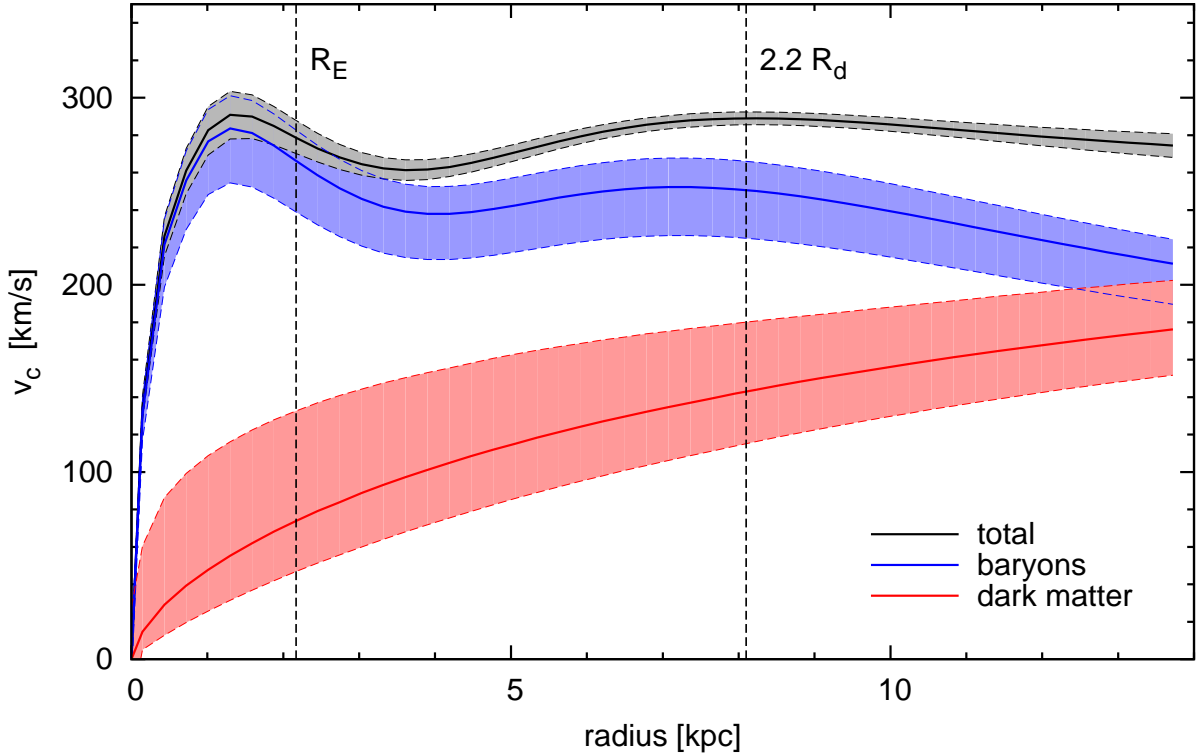}}
  \caption{Circular velocity profile inferred from the combined
    lensing and dynamics analysis. The total circular velocity is
    shown in black, while the baryonic and dark matter components are
    plotted in blue and in red, respectively. For each component, the
    solid line represents the median while the shaded region encloses
    68\% of the posterior PDF. The two vertical dashed lines mark, for
    reference, the location of the Einstein radius $\REin$ (left) and
    of $2.2 \Rd$ (right), where $\Rd = 3.58$ kpc is the disk scale
    length of the galaxy.}
  \label{fig:prof_vc}
\end{figure*}
% -------------------------------------------------------------------

\subsection{Mass budget: baryons and dark matter}
\label{ssec:fDM}

A very intuitive way to visualize the galaxy mass budget as a function
of radius that is inferred from the combined analysis is provided by
\Fref{fig:prof_vc}, where we show the circular velocity profile
obtained from the posterior PDF, decomposed into the baryonic and dark
matter components. The solid lines indicate the median values from the
posterior PDF, while the shaded regions represent the 68 per cent
confidence intervals. 
The constraints on the total circular velocity $\vtot$ are extremely
tight, whereas there are larger uncertainties on the contributions
given by the separate components. Despite this, it is clear that the
baryonic matter is dominant all over the entire region for which we
have data, with the dark matter component becoming progressively more
important as we move outwards in radius.

\begin{figure}
  \centering
  \resizebox{1.00\hsize}{!}{\includegraphics[angle=-90]
            {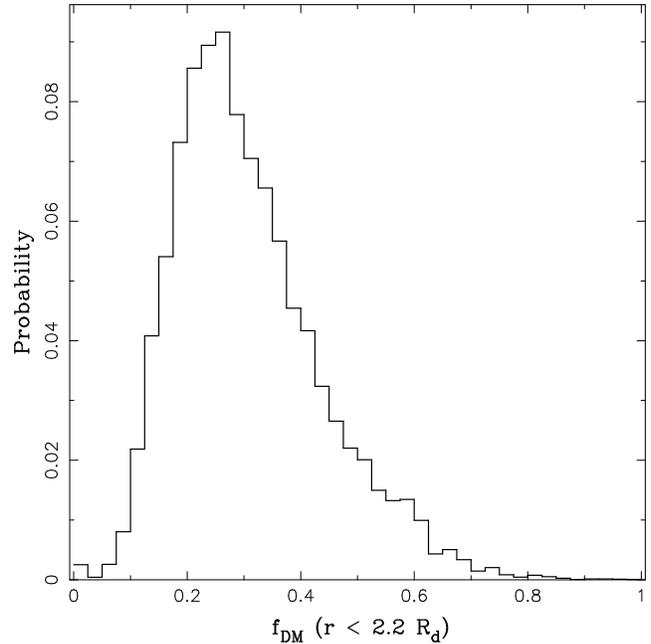}}
  \caption{Dark matter fraction enclosed within the spherical radius
    $r = 2.2$ disk scale lengths, inferred from the combined lensing
    and dynamics analysis. The median and uncertainty (corresponding
    to 16th and 84th percentiles) is $\fDM = 0.28_{-0.10}^{+0.15}$.}
  \label{fig:fDM}
\end{figure}

Traditionally, in studies of disk galaxies, the characteristic radius
at which one measures the dark matter fraction~$\fDM \equiv
\MDM/\Mtot$ is 2.2~times the disk scale length~$\Rd$, which
corresponds to the radius at which the circular speed peaks for a
razor-thin exponential disk (see, e.g., \citealt{Bershady2010} and
references therein). In the case of {\jdisk}, we determine
$\fDM(2.2\Rd) = 0.28^{+0.15}_{-0.10}$, by integrating the mass within
a spherical radius $r = 2.2\Rd$. We note that this dark matter
fraction was inferred from gravitational data alone, and is
independent of the stellar populations in the galaxy.  The
marginalized posterior PDF for this quantity is shown in
\Fref{fig:fDM}: it is clear that the distribution peaks at around
$\fDM \simeq 0.3$; dark matter dominated models (i.e. $\fDM > 0.5$),
however, are still possible, albeit with a low probability of about~9
percent. Models without dark matter, on the other hand, are ruled out
at more than the 3-sigma level, i.e. the probability for $\fDM < 0.05$
is less than $0.3$~per cent. The Paper~II analysis of this same
galaxy\,---\,carried out using a less flexible mass model and without
including the stellar kinematic constraints\,---\,found (at lower
precision) a higher contribution of dark matter at $2.2\Rd$,
i.e. $\fDM = 0.55^{+0.20}_{-0.15}$, which is however still consistent
within 1-sigma with the result that we determine
here. \citet{Trott2010} and \citet{Suyu2011}, by applying a combined
lensing and dynamics analysis on two different disk galaxies, obtain a
fractional amount of dark matter close to~$45\pm10$ percent, which is
slightly higher than (but not inconsistent with) the $\fDM$ that we
derive from the present analysis. On the other hand,
\citet{vandeVen2010}, by conducting a combined lensing and dynamics
study of the same early-type disk galaxy studied by \citet{Trott2010},
and adopting a \citet{Kroupa2001} IMF, find that the upper limit for
$\fDM$ is only $\approx 0.20$. Interestingly, the value of $\fDM$ that
we obtain in this analysis is similar to the typical average dark
matter fraction of about 30~per cent determined for massive early-type
galaxies within one effective radius based on lensing and dynamical
analysis \citep[e.g.,][]{Treu-Koopmans2004, Treu2010, Auger2010imf,
  Spiniello2011}, or by assuming maximal stellar component (e.g.,
\citealt{Gerhard2001}, \citealt{Cappellari2006},
\citealt{Barnabe2010}, \citeyear{Barnabe2011}).  However, one should
keep in mind that $\fDM$ has been observed to vary quite significantly
between individual systems.

This analysis also enables us to test whether the ``maximum disk''
hypothesis \citep{vanAlbada-Sancisi1986}, frequently adopted in
studies of late-type galaxies \citep[e.g.][]{Bell-deJong2001}, holds
for the object examined here. We follow the definition of maximum disk
introduced by \citet{Sackett1997}, i.e. $\vdisk(2.2 \Rd) / \vtot(2.2
\Rd) = 0.85 \pm 0.10$, substituting the circular velocity of the
disk~$\vdisk$ with the more relevant circular velocity of the entire
baryonic component, $\vbar$. We find that $\vbar(2.2 \Rd) / \vtot(2.2
\Rd) = 0.87^{+0.05}_{-0.09}$, which corresponds to a maximal
disk. From the posterior PDF for this ratio, the probability that the
{\jdisk} disk is submaximal is about $10$ per cent.

This result would make {\jdisk} something of an outlier when compared
with a sample, recently studied using dynamical methods, of 30 local
disk galaxies \citep{Bershady2011, Martinsson2011PhD}. These authors
find that, although the ratio $\vdisk(2.2 \Rd) / \vtot(2.2 \Rd)$
increases with the maximum rotation speed of the galaxy, even the most
massive systems with $\vdisk(2.2 \Rd) \gtrsim 250$ km s$^{-1}$ are
submaximal on average. We note, however, that the existence of
individual massive galaxies consistent with ``maximality'' is not
ruled out in their study (see, in Figure~2 of \citealt{Bershady2011},
the outlier and the error bars for some of the highest rotation
velocity systems). In addition, it is important to keep mind that both
our method and that of \citet{Bershady2011} inevitably rely on
different assumptions: in our case, for example, a common
mass-to-light ratio for bulge and disk, in their case assumptions
necessary to compare edge-on and face-on galaxies. In addition, the
methods obtain their information from different parts of the mass
distribution, with our method being more sensitive to the inner
regions, owing to the lensing and stellar velocity dispersion
constraints. We plan to perform a more detailed comparison of the two
results once data and models for the full SWELLS sample will be
available.

 %  DM halo  %  %  %  %  %  %  %  %  %  %  %  %  %  %  %  %  %  %  % 

\subsection{Constraints on the dark matter halo: shape and profile}
\label{ssec:halo}

Pure dark matter N-body simulations find that dark halos generally
have triaxial shapes, with a preference for prolateness, particularly
in the inner regions
\citep[e.g][]{Jing-Suto2002,Allgood2006,Bett2007,Maccio2008}.  Recent
numerical work \citep[see, e.g.,][]{Abadi2010,Tissera2010} has shown
that including the contribution of the baryons has the effect of
modifying the overall profile of the dark halo, which flattens to a
more axisymmetric and oblate shape, with an average axial ratio of
order $0.85 - 0.95$, largely constant with radius. In our study of
{\jdisk}, we infer from the combined analysis that the dark halo is
moderately oblate, with an axial ratio~$\qh = 0.75^{+0.27}_{-0.16}$.
Significantly prolate haloes with $\qh > 1.25$ are strongly
disfavoured (i.e., with less than 5 per cent probability). This is in
good agreement with the numerical results on baryon-affected halos,
although the median value is slightly flatter than the typical $\qh$
obtained in the simulations. The axial ratio obtained for {\jdisk} in
Paper~II (using a less flexible NIE dark halo model which does not
allow for a variable inner slope), i.e. $\qh = 0.91^{+0.15}_{-0.13}$,
was closer to spherical but still consistent, within the 68\%
uncertainty, with the more accurate analysis conducted here. In
contrast, in the only other joint lensing and kinematics study of a
disk galaxy that adopts a non-spherical halo model, \citet{Suyu2011}
find a much more flattened dark matter distribution, with $\qh =
0.33$. These authors adopt a simpler luminous mass model than the one
considered here (i.e., a razor-thin exponential disk and a point-mass
bulge) and do not have access to stellar kinematics data.

Including the inner slope~$\slope$ as a free parameter in the dark
halo mass density model (see \Sref{ssec:mass-model:DM}), rather than
just adopting a fixed isothermal or NFW profile (as done in previous
studies, cf.\ Paper~II and \citealt{Suyu2011}; but see also
\citealt{Trott2010}, where a spherical gNFW halo is used), is
important since it allows one to account for possible baryon-induced
effects, such as adiabatic contraction, that can modify the steepness
of the density distribution in the galaxy central regions. The
data-set at hand, unfortunately, does not permit us to place a strong
constraint on the inner slope: we obtain $\slope =
0.82^{+0.65}_{-0.54}$, approximately equiprobable over the range 0 to
1.5, and perfectly consistent with an unmodified NFW profile.  The
probability that the halo has an inner slope $1.5 \le \slope < 2$ is
14 per cent. We are able to conclude, however, that very steep
profiles are disfavoured: slopes $1.7 \le \slope < 2$ have only a
$3.5$ per cent probability, whereas from the adopted uniform prior
U(0,2) (see~\Tref{table:par}) one would predict 15 per cent over the
same interval.

We infer a halo concentration parameter $\cg = 7.7_{-2.5}^{+4.2}$,
with a low-probability tail for high concentrations (the 95th and 98th
percentiles fall at $\cg \simeq 17$ and $\cg \simeq 30$,
respectively). The inferred virial velocity is $\vvir =
242_{-39}^{+44}$. From these parameters one can derive the posterior
PDFs for all other useful quantities characterizing the gNFW halo,
such as the generalized scale radius $\rsg = 41_{-19}^{+27}$ kpc, the
virial radius $\rvir = 315_{-53}^{+57}$ kpc and the virial mass
$\log_{10} (\Mvir/M_{\sun}) = 12.48_{-0.27}^{+0.28}$.

\begin{figure}
  \centering
  \resizebox{1.00\hsize}{!}{\includegraphics[angle=0]
            {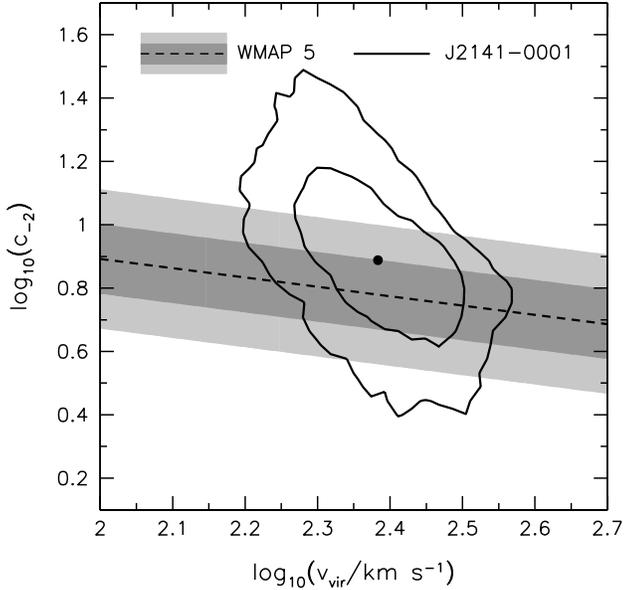}}
          \caption{Dark matter concentration $\cg \equiv \rvir/\rsg$
            vs dark matter virial velocity $\vvir$. The shaded region
            shows the prediction (with 1- and 2-sigma scatter) from
            N-body simulations in a WMAP 5th year $\Lambda$CDM
            cosmology \citep{Maccio2008}. The contours enclose 68\%
            and 95\% of the posterior probability from the combined
            lensing and dynamics analysis. The black dot shows the
            median of the posterior distribution.}
  \label{fig:cDM}
\end{figure}

\Fref{fig:cDM} shows a comparison between the dark matter
concentration and the virial velocity from our lensing and dynamics
analysis of {\jdisk} (contours), with the predictions from N-body
simulations in a WMAP 5th year cosmology \citep{Maccio2008}.  The
uncertainty on our inferred dark matter concentration is quite broad
$\simeq 0.2$ dex, but is nevertheless in very good agreement with the
simplest predictions from $\Lambda$CDM (i.e., assuming no contraction
or expansion of the dark matter in response to galaxy formation).

% -------------------------- DM FRACTION ----------------------------
% --------------------------------------------------------------------

% -------------------------- DM CONCENTRATION ------------------------
% --------------------------------------------------------------------

 %  IMF  %  %  %  %  %  %  %  %  %  %  %  %  %  %  %  %  %  %  %  %  % 

\subsection{Constraints on the stellar IMF}
\label{ssec:IMF}

The total stellar mass inferred from the combined lensing and dynamics
analysis is $\log_{10} (\Ms/\Msun) = 11.12_{-0.09}^{+0.05}$. This
value is very well constrained and represents a significant
improvement over the $\Ms$ determination of Paper~II, by cutting the
low mass tail of the posterior PDF of about $0.3$~dex.

In order to draw conclusions on the galaxy IMF, we need to compare the
stellar mass derived from the joint analysis with the stellar masses
that are inferred from SPS models when assuming either a
\citet{Chabrier2003} or a \citet{Salpeter1955} IMF. The SPS analysis
is performed by applying the \citet{Auger2009} code on the multi-band
photometric data set of {\jdisk}, as described in Paper~II.  However,
we note that so far we have neglected the contribution of the cold
gas: if such a component is present, the mass $\Ms$ derived above from
the combined analysis actually represents the total \emph{baryonic}
mass. Therefore, in order to obtain a posterior PDF for the stellar
mass that can be properly compared with the predictions from the SPS
models, we need to subtract the cold gas fraction, which in disk
galaxies (with stellar masses of $\Ms\simeq 10^{11}\Msun$) accounts
for about $20 \pm 10$ per cent of the baryonic mass \citep[see
  e.g.][]{Dutton-vandenBosch2009}.  Under the assumption that the cold
gas is distributed approximately like the stars, for each sample in
the posterior PDF of $\Ms$ we draw a random gas fraction
$f_{\mathrm{gas}} \in [0,1]$ from a Gaussian distribution centered on
0.2 with a standard deviation of 0.1, and we calculate the quantity
$\Ms (1-f_{\mathrm{gas}})$. The gas-subtracted stellar mass derived in
this way is $\log_{10} (\Ms/\Msun) = 11.01_{-0.11}^{+0.08}$, about 0.1
dex lower than the value obtained above by ignoring the cold gas
contribution. This provides a robust lower bound to the stellar mass
function. In the future, it will be useful to refine these mass
estimates by including high resolution constraints on the gas fraction
from ALMA observations.

\begin{figure*}
  \centering
  \resizebox{1.00\hsize}{!}{\includegraphics[angle=-90]
            {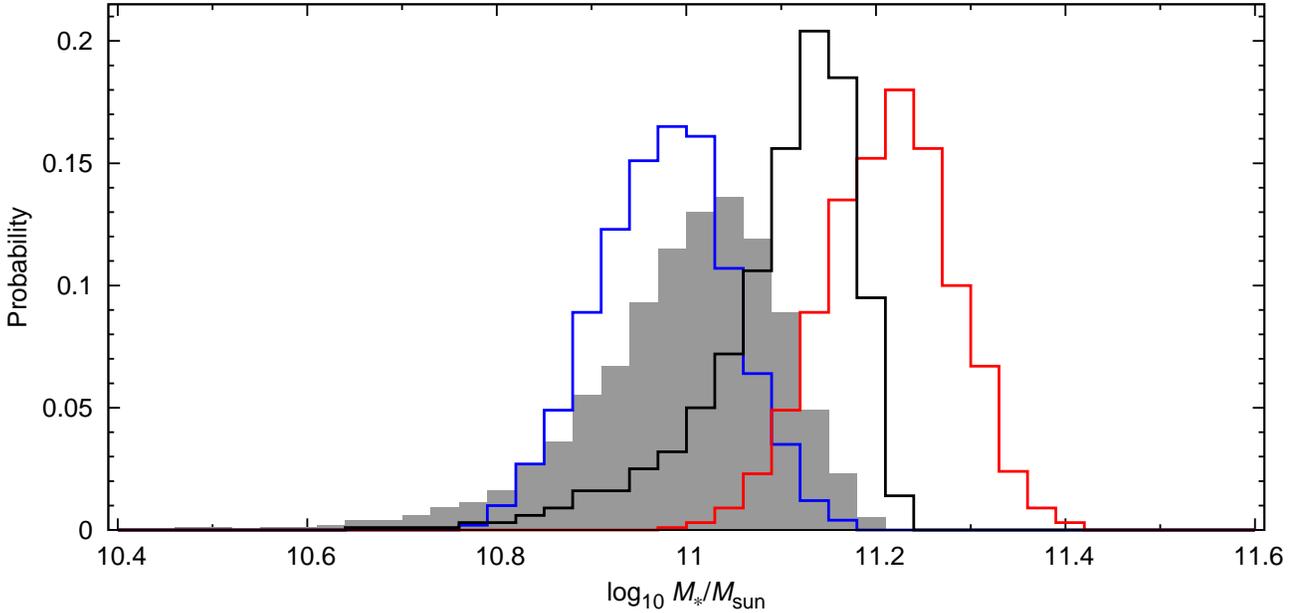}}
  \caption{Comparison of the stellar mass inferred from the combined
    lensing and dynamics analysis (black) with the stellar mass
    determined from photometry and stellar population synthesis
    models, assuming a Chabrier IMF (blue) or a Salpeter IMF
    (red). The grey shaded histogram shows the posterior PDF for the
    inferred stellar mass when a 20\% $\pm$ 10\% contribution in mass
    from cold gas in included. In the latter case, the Bayes factor in
    favor of a Chabrier IMF with respect to the Salpeter IMF is 5.7,
    corresponding to Chabrier being preferred at 85\%.}
  \label{fig:IMF}
\end{figure*}

The posterior PDF for the inferred stellar mass (both with and without
cold gas) is presented in \Fref{fig:IMF}, and compared with the
distributions obtained for a Chabrier and Salpeter IMF. It is clear,
just by a visual inspection of this Figure, that our results support a
Chabrier-like IMF over a Salpeter one. This preference can be
quantified in a rigorous way by calculating the Bayes factor, i.e. the
evidence ratio between the two models
\begin{equation}
  \label{eq:bayes-factor}
  B_{CS} = 
  \frac{\int \mathcal{L}(\Ms) \, \pr(\Ms \, | \, \mathcal{H}_{C}) \, \de \Ms}
       {\int \mathcal{L}(\Ms) \, \pr(\Ms \, | \, \mathcal{H}_{S}) \, \de \Ms} 
       \, ,
\end{equation}
where in our case the likelihood $\mathcal{L}(\Ms)$ is the posterior
PDF for the inferred stellar mass, while the priors $\pr(\Ms \, | \,
\mathcal{H}_{C})$ and $\pr(\Ms \, | \, \mathcal{H}_{S})$ are given by
the posterior PDFs obtained from SPS models in the cases of Chabrier
and Salpeter IMFs, respectively. The calculated Bayes factor is
$B_{CS} = 5.7$, which corresponds to substantial evidence in favour of
a Chabrier IMF with respect to a Salpeter IMF \citep[see e.g.][and
  references therein]{Kass-Raftery1995}. In other words, if these are
the only two possible models, this value of $B_{CS}$ means that there
is a 85 per cent probability that the Chabrier model is the true one.

% -------------------------- IMF COMPARISON -------------------------
% --------------------------------------------------------------------

This result corroborates the findings of Paper~II, and is in agreement
with the works of, e.g, \citet{Bell-deJong2001}, \citet{Kassin2006}
and \citet{vandeVen2010}, which disfavour a Salpeter IMF for disk
galaxies, preferring instead IMFs that predict lower stellar masses,
such as Chabrier or \citet{Kroupa2001}. Low-mass, fast-rotating
early-type galaxies are also found to be often inconsistent with a
Salpeter IMF \citep[e.g.][]{Cappellari2006, Auger2010,
  Barnabe2010}. On the other hand, there is mounting evidence that
massive ellipticals favour a Salpeter-like \citep{Auger2010imf,
  Treu2010, Barnabe2011, Spiniello2011} or an even steeper IMF
\citep{vanDokkum-Conroy2010}. These findings, including the results of
this work, support the idea that the traditional picture of a
universal IMF \citep[see, e.g.,][]{Kroupa2002} might need to be
revised in favour of a more complicated scenario where the IMF depends
on the galaxy mass and/or Hubble type.

 %  Anisotropy  %  %  %  %  %  %  %  %  %  %  %  %  %  %  %  %  %  % 

\subsection{Constraints on the stellar anisotropy}
\label{ssec:beta}

Determining the shape of the velocity dispersion ellipsoid of disk
galaxies is important not only in order to understand their global
dynamical properties, which are related to the formation and evolution
mechanisms, but also because the vertical-to-radial velocity
dispersion ratio $\sigma_{z}/\sigma_{R}$ can be used, together with
the galaxy scale height distribution, to derive the dynamical mass of
the disk \citep[see e.g.][]{Bottema1997, Kregel2005, Westfall2011}.

From our analysis, we infer a meridional anisotropy parameter $b =
1.77_{-0.29}^{+0.30}$, with a very symmetric posterior distribution
around the median value.  In order to facilitate the comparison with
the disk galaxy studies literature, it is convenient to express the
anisotropy in the notation $\beta_{z} = 1 -
\sigma^{2}_{z}/\sigma^{2}_{R}$ (see Sect.~\ref{ssec:jeans}), where
$\beta_{z} = 0$ corresponds to isotropy: in this case we have
$\beta_{z} = 0.43_{-0.11}^{+0.08}$.  These results show that,
for~{\jdisk}, the velocity dispersion in the vertical direction is
about three quarters of the radial velocity dispersion. The
probability that the velocity dispersion ellipsoid is approximately
isotropic in the meridional plane (i.e., $\beta_{z} = 0.0 \pm 0.1$) is
only of order 1~per cent. This confirms that two-integral DF models
(which are semi-isotropic, i.e. have $\sigma^{2}_{R} = \sigma^{2}_{z}$
everywhere, see e.g. \citealt{BT08}) do not provide an ideal
description of the dynamical properties of this galaxy, and a more
flexible approach allowing for anisotropy, such as the one adopted in
this work, is warranted. Within the hypothesis of axial symmetry, we
can then conclude that the disk galaxy DF respects a third integral of
motion \citep[cf., e.g.,][]{Noordermeer2008}.

These findings are in agreement with numerous dynamical studies of
disk galaxies, including the Milky Way, which are well known to have
velocity dispersion ellipsoids flattened along the vertical direction
\citep[see][and references therein]{vanderKruit-Freeman2011}.  For
local disk galaxies, \citet{Gerssen1997, Gerssen2000} and
\citet{Shapiro2003} determine $0.30 \lesssim \beta_{z} \lesssim
0.75$. \citet{Williams2009}, adopting a dynamical model analogous to
the one used in this work (i.e., based on anisotropic Jeans
equations), find $0.0 \lesssim \beta_{z} \lesssim 0.5$ for a sample
of~14 spiral and S0 galaxies. \citet{Noordermeer2008}, using
two-dimensional kinematic data sets to analyze the dynamics of four
early-type disk galaxies, find $\beta_{z} \simeq 0.5$, perfectly
consistent with the result for {\jdisk}. Recently, one of the galaxies
studied in detail in the DiskMass Survey was determined to have a more
flattened $\beta_{z} = 0.77$ \citep{Westfall2011}.

We remark that the present study represents the first determination of
the anisotropy parameter for a disk galaxy well beyond the local
Universe, at a redshift $\zL \simeq 0.14$ (a previous combined lensing
and dynamics study of a disk galaxy at a lower redshift, $\zL \simeq
0.04$, was conducted by \citealt{vandeVen2010}, who found $\beta_{z} =
0.1\pm0.1$, consistent with the system being semi-isotropic).

%%%%%%%%%%%%%%%%%%%%%%%%%%% CONCLUSIONS %%%%%%%%%%%%%%%%%%%%%%%%%%%%%%

\section{Conclusions}
\label{sec:conclude}

We have carried out an in-depth, self-consistent analysis of the mass
and dynamical structure of the lens disk galaxy {\jdisk} at
redshift~$0.14$ by combining the constraints from gravitational
lensing, H$\alpha$ rotation curve and stellar kinematics. We have
adopted a flexible axially symmetric mass model consisting of a gNFW
dark matter halo and a self-gravitating stellar distribution obtained
from the MGE parametrization of the observed luminous profile. We have
modelled the kinematics by means of anisotropic Jeans equations in
order to allow for a velocity dispersion ellipsoid that is flattened
in the meridional plane, as is typical for disk galaxies. 

This work improves in several ways (namely, the inclusion of stellar
kinematics constraints and the upgraded mass and dynamical model) on
the study of this same object described in Paper~II, and represents
the most accurate and detailed analysis to date of the dark and
luminous mass profile of a disk galaxy beyond the local Universe,
i.e. at redshift $\gtrsim 0.1$. The main conclusions of this analysis can
be summarized as follows:
\begin{enumerate}
\item The spherical dark matter mass fraction within $2.2 \Rd$ is
  $\fDM = 0.28^{+0.15}_{-0.10}$, independent of assumptions on the
  stellar populations in the galaxy. The dark matter fraction
  increases with radius, but does not become dominant within the range
  probed by the observations, which extend to approximately $R = 14$
  kpc. Models without dark matter (i.e., $\fDM < 0.05$) are ruled out
  at more than the 3-sigma level.
\item We test the maximum disk hypothesis: we find that, at $2.2 \Rd$,
  the fractional contribution of the baryons to the total circular
  velocity is $0.87^{+0.05}_{-0.09}$. This corresponds to a maximal
  disk (following the definition of \citealt{Sackett1997}); the
  probability of having a submaximal disk for {\jdisk} is 10 per
  cent. This is in disagreement with recent studies of local disk
  galaxies (e.g., the DiskMass Survey, \citealt{Bershady2011},
  \citealt{Martinsson2011PhD}), which typically find submaximal disks.
\item The gNFW dark matter halo is characterized by a virial velocity
  $\vvir = 242_{-39}^{+44}$ $\kms$ and a concentration parameter $\cg =
  7.7_{-2.5}^{+4.2}$, implying a generalized scale radius $\rsg =
  41_{-19}^{+27}$ kpc. This is in very good agreement with the
  predictions from N-body simulations in a $\Lambda$CDM cosmology
  (i.e., assuming no contraction or expansion of the halo in response
  to galaxy formation).
\item The inner slope of the dark matter halo is only weakly
  constrained, $\slope = 0.82^{+0.65}_{-0.54}$, and is consistent with
  an unmodified NFW profile ($\slope = 1$). We can still conclude,
  however, that very steep inner profiles with $\slope \gtrsim 1.7$
  are disfavoured.
\item The dark matter halo is moderately oblate, with a
  three-dimensional axial ratio $\qh = 0.75^{+0.27}_{-0.16}$, and a
  very low probability for significantly prolate haloes (i.e., $\qh
  \gtrsim 1.25$). Recent high-resolution simulations
  \citep[e.g.][]{Abadi2010, Tissera2010} find that the baryons have
  the effect of turning the prolate triaxial dark matter halos into
  roughly oblate spheroids, a scenario that is consistent with the
  results of this work.
\item The total baryonic mass is tightly constrained by the combined
  lensing and dynamics analysis, and is determined to be $\log_{10}
  (\Ms/\Msun) = 11.12_{-0.09}^{+0.05}$, independent of the IMF. When
  accounting for the expected cold gas contribution, we obtain a
  stellar mass $\log_{10} (\Ms/\Msun) = 11.01_{-0.11}^{+0.08}$. This
  value is in excellent agreement with the stellar mass that is
  predicted when assuming a Chabrier IMF. Model comparison shows that
  there is substantial evidence in favour of a Chabrier IMF with
  respect to a Salpeter IMF (the Bayes factor is 5.7, corresponding to
  a 85 per cent probability).
\item We infer a meridional anisotropy parameter $\beta_{z} =
  0.43_{-0.11}^{+0.08}$, implying that, for {\jdisk}, the velocity
  dispersion ellipsoid in the meridional plane is flattened along the
  vertical direction, in agreement with most studies of local disk
  galaxies. Semi-isotropic models (i.e., $\beta_{z} \approx 0$) are
  ruled out at a very high confidence level, corroborating the
  evidence that the dynamics of disk galaxies is not adequately
  described by two-integral DFs, and a third integral of motion is
  required.
\end{enumerate}

%%%%%%%%%%%%%%%%%%%%%%%%%%% ACKNOWLEDGMENTS %%%%%%%%%%%%%%%%%%%%%%%%%%

\section*{Acknowledgments}

%% We thank XXX for useful discussions.
%
MB acknowledges support from the Department of Energy contract
DE-AC02-76SF00515.
AAD acknowledges financial support from a CITA National Fellowship,
from the National Science Foundation Science and Technology Center
CfAO, managed by UC Santa Cruz under cooperative agreement
No. AST-9876783. AAD and DCK were partially supported by NSF grant AST
08-08133, and by HST grants AR-10664.01-A, HST AR-10965.02-A, and HST
GO-11206.02-A.
PJM was given support by the TABASGO and Kavli foundations, and the
Royal Society, in the form of research fellowships.
TT acknowledges support from the NSF through CAREER award NSF-0642621,
and from the Packard Foundation through a Packard Research Fellowship.
LVEK acknowledges the support by an NWO-VIDI programme subsidy
(programme number 639.042.505).
This research is supported by NASA through Hubble Space Telescope
programs GO-10587, GO-11202, and GO-11978, and in part by the National
Science Foundation under Grant No. PHY99-07949. and is based on
observations made with the NASA/ESA Hubble Space Telescope and
obtained at the Space Telescope Science Institute, which is operated
by the Association of Universities for Research in Astronomy, Inc.,
under NASA contract NAS 5-26555, and at the W.M. Keck Observatory,
which is operated as a scientific partnership among the California
Institute of Technology, the University of California and the National
Aeronautics and Space Administration. The Observatory was made
possible by the generous financial support of the W.M. Keck
Foundation. The authors wish to recognize and acknowledge the very
significant cultural role and reverence that the summit of Mauna Kea
has always had within the indigenous Hawaiian community.  We are most
fortunate to have the opportunity to conduct observations from this
mountain.
Funding for the SDSS and SDSS-II was provided by the Alfred P. Sloan
Foundation, the Participating Institutions, the National Science
Foundation, the U.S. Department of Energy, the National Aeronautics
and Space Administration, the Japanese Monbukagakusho, the Max Planck
Society, and the Higher Education Funding Council for England. The
SDSS was managed by the Astrophysical Research Consortium for the
Participating Institutions. The SDSS Web Site is http://www.sdss.org/.

%%%%%%%%%%%%%%%%%%%%%%%%%%%%% BIBLIOGRAPHY %%%%%%%%%%%%%%%%%%%%%%%%%%%

%\bibliography{../../DYNLEN/OURPAPERS/my_bibliography}

\label{lastpage}

\clearpage

\end{document}